\documentclass{aa}  

\pdfoutput=1
\usepackage{longtable}
\usepackage{natbib,afterpage}
\usepackage{color}
\usepackage{amssymb,times}
\usepackage[english]{babel}
\usepackage{subfigure}
\usepackage{ifpdf}
\usepackage{morefloats}
\usepackage[rightcaption]{sidecap}

\newif\ifpdf \ifx\pdfoutput\undefined\pdffalse\else\pdftrue\fi
\ifpdf\pdfoutput=1
        \usepackage{graphicx}
        \usepackage{epstopdf}
        \DeclareGraphicsExtensions{.pdf,.png,.gif,.jpg}
        \else \usepackage[dvips]{color,graphicx} \fi

\def\Msun{\hbox{M$_{\odot}$}}               
\def\Rsun{\hbox{R$_{\odot}$}}               
\def\Lsun{\hbox{L$_{\odot}$}}               
\def\Rstar{\hbox{R$_{\star}$}}              
\def\Mdot{\hbox{$\dot{M}$}}               
\def\arcsec{\hbox{$^{\prime\prime}$}}
\def\arcmin{\hbox{$^{\prime}$}}
\def\deg{\hbox{$^\circ$}}       

\begin{document} 

\selectlanguage{english}
\newcommand{\red}{\textcolor[rgb]{1,0,0}}
\newcommand{\blue}{\textcolor[rgb]{0,0,1}}

\newlength{\fntxvi} \newlength{\fntxvii}
\newcommand{\chemical}[1]
{{\fontencoding{OMS}\fontfamily{cmsy}\selectfont
\fntxvi\the\fontdimen16\font
\fntxvii\the\fontdimen17\font
\fontdimen16\font=3pt\fontdimen17\font=3pt
$\mathrm{#1}$
\fontencoding{OMS}\fontfamily{cmys}\selectfont
\fontdimen16\font=\fntxvi \fontdimen17\font=\fntxvii}}


\title{ALMA-resolved salt emission traces the chemical footprint and inner wind morphology of VY~CMa}

   \author{
   L. Decin\inst{1}
          \and
          A. M. S. Richards\inst{2}
          \and
          T. J. Millar\inst{3}
          \and
          A. Baudry\inst{4,5}
          \and
          E. De Beck\inst{6}
          \and
          W. Homan\inst{1}
          \and
          N. Smith\inst{7}
          \and
          M. Van de Sande\inst{1}
          \and
          C. Walsh\inst{8}}

  \offprints{Leen.Decin@ster.kuleuven.be}

  \institute{
  Instituut voor Sterrenkunde, Katholieke Universiteit Leuven, Celestijnenlaan 200D, 3001 Leuven, Belgium  
  \email{Leen.Decin@ster.kuleuven.be}
  \and
 JBCA, Department Physics and Astronomy, University of Manchester, Manchester M13 9PL, UK
 \and
 Astrophysics Research Centre, School of Mathematics and Physics, Queen's University Belfast, University Road, Belfast, BT7 1NN, UK
 \and
 Universit\'e de Bordeaux, LAB, UMR 5804, 33270 Floirac, France
 \and
 CNRS, LAB, UMR 5804, 33270 Floirac, France
 \and
 Department of Earth and Space Sciences, Chalmers University of Technology, Onsala Space Observatory, 43992 Onsala, Sweden
 \and
 Steward Observatory, University of Arizona, 933 N. Cherry Ave, Tucson, AZ 85721, USA
 \and
 Leiden Observatory, Leiden University, PO Box 9513, 2300 RA, Leiden, The Netherlands
}


 
  \abstract
   {At the end of their lives, most stars lose a significant amount of mass through a stellar wind. The specific physical and chemical circumstances that lead to the onset of the stellar wind for cool luminous stars are not yet understood. Complex geometrical morphologies in the circumstellar envelopes prove that various dynamical and chemical processes are interlocked and their relative contributions are not easy to disentangle.}
   {We aim to study the inner-wind structure ($R<250$\,\Rstar) of the well-known red supergiant VY~CMa, the archetype for the class of luminous red supergiant stars suffering high mass loss. Specifically, the objective is to unravel the density structure in the inner envelope and to examine the chemical interaction between some gas and dust species.  }
   {We analyse high spatial resolution ($\sim 0\farcs24\times0\farcs13$) ALMA Science Verification (SV) data in band~7 in which four thermal emission lines of gaseous sodium chloride (NaCl) are present at high signal-to-noise ratio.}
   {For the first time, the NaCl emission in the inner wind region of VY~CMa is spatially resolved. The ALMA observations reveal the contribution of up to four different spatial regions. The NaCl emission pattern is different compared to the dust continuum and TiO$_2$ emission already analysed from the ALMA SV data. The emission can be reconciled with an axisymmetric geometry, where the lower density polar/rotation axis has a position angle of $\sim$50\deg\ measured from north to east. However, this picture can not capture the full morphological diversity, and discrete mass ejection events need to be invoked to explain localized higher-density regions. The velocity traced by the gaseous NaCl line profiles is significantly lower than the average wind terminal velocity, and much slower than some of the fastest mass ejections, signalling a wide range of characteristic speeds for the mass loss. Gaseous NaCl is detected far beyond the main dust condensation region. Realising the refractory nature of this metal halide, this hints at a chemical process preventing all NaCl from condensing onto dust grains. We show that in the case of the ratio of the surface binding temperature to the grain temperature being $\sim$50, only some 10\% of NaCl remains in gaseous form, while for lower values of this ratio thermal desorption efficiently evaporates NaCl.  Photodesorption by stellar photons seems not to be a viable explanation for the detection of gaseous NaCl at 220\,\Rstar\ from the central star, and instead, we propose shock-induced sputtering driven by localized mass ejection events as alternative.}
   {The analysis of the NaCl lines demonstrate the capabilities of ALMA to decode the geometric morphologies and chemical pathways prevailing in the winds of evolved stars. These early ALMA results prove that the envelopes surrounding evolved stars are far from homogeneous, and that a variety of dynamical and chemical processes dictate the wind structure.}

   
\titlerunning{ALMA-resolved salt emission in VY~CMa}
  \maketitle

\section{Introduction}\label{Sec:introduction}

The post-main-sequence evolution of stars with an initial mass in the range of 20-40\,\Msun\ may lead through a cool red supergiant (RSG) phase with strong mass loss. These stars can potentially evolve through a Wolf-Rayet phase as well if mass loss is sufficient to remove the hydrogen envelope, and if not, they may explode as a supernova \citep{Smith2009AJ....137.3558S}. The amount and nature of the mass loss dictates the future evolution of the star and determines their supernova (SN) appearance, be it as SN of type II-P, II-L, IIb, IIn, or Ib/c. If a dense circumstellar envelope is still present at the moment of explosion, it will provide an obstacle for the SN blast wave to overtake \citep{Pun2002ApJ...572..906P}.

At the present time the driving mechanism of the mass loss of RSGs is not yet understood. Mechanisms based on radiation pressure on freshly synthesized dust grains, magneto-acoustic waves, and turbulent pressure due to convection in combination with radiative pressure on molecular lines \citep{Josselin2007A&A...469..671J} are invoked, or a combination of the above in a so-called hybrid model featuring magneto-rotational effects with a dust driven-wind scenario \citep{Thirumalai2012MNRAS.422.1272T}. Not being able to constrain the driving mechanism(s) represents a serious limitation for current stellar evolution models which therefore adopt a simplified parametrized, time averaged, mass-loss rate prescription \citep[e.g.][]{Reimers1975MSRSL...8..369R, deJager1988A&AS...72..259D, Nieuwenhuijzen1990A&A...231..134N}. Empirical studies, however, show that the mass loss rate, $\Mdot(t)$, may vary significantly during the RSG evolution \citep[e.g.,][]{Decin2006A&A...456..549D} and that these stars can display sudden episodic bursts of mass loss potentially linked to the magnetic and/or convective nature of the star \citep{Smith2001AJ....121.1111S, Adande2013ApJ...778...22A}. 

The circumstellar envelopes (CSEs) created by the intense mass loss are rich chemical laboratories in the universe. Because of the low temperatures in the envelopes and the long duration of the mass loss, molecules and solid-state species are formed in these CSEs. Through stellar winds, these elements are injected into the interstellar medium (ISM) thereby gradually enriching their surroundings.

Efforts to deepen our understanding of the origin of the mass-loss and of the thermodynamical and chemical behaviour of the CSE proceed on two separate tracks: more detailed studies of specific RSGs which are often selected by virtue of their proximity, and global studies of a larger sample of RSGs to derive general characteristics. The first of these approaches is pursued in this paper in which we present ALMA high-spatial resolution ($\sim0\farcs24\times0\farcs13$) observations of the close-by RSG VY~CMa in thermal sodium chloride (NaCl) emission lines tracing the inner wind region within $\sim$1.5\arcsec\ (or $\sim$280\,\Rstar) from the central star. 

\subsection{The red supergiant VY CMa}
VY~CMa is one of the most intrinsically luminous red stars in the galaxy \citep[$L\sim3\times10^5$\,\Lsun;][]{Smith2001AJ....121.1111S, Choi2008PASJ...60.1007C}. Its current mass is estimated around 17\,\Msun, but initially it was some 25\,\Msun\ \citep{Wittkowski2012A&A...540L..12W}. VY~CMa is also known to be one of the largest RSGs with a stellar radius, \Rstar, of 1420$\pm$120\,\Rsun\ \citep{Wittkowski2012A&A...540L..12W} or 5.53\,mas at a distance of 1.2$\pm$0.1\,kpc \citep{Choi2008PASJ...60.1007C, Zhang2012ApJ...744...23Z}. The average recent mass-loss rate is high, around $1\!-\!4\times10^{-4}$\,\Msun\,yr$^{-1}$ \citep{Danchi1994AJ....107.1469D, Monnier1999ApJ...512..351M, Decin2006A&A...456..549D}, producing a large circumstellar nebula some 10\arcsec\ (or 0.06\,pc) across \citep{Humphreys2007AJ....133.2716H, Muller2007ApJ...656.1109M}. The stellar velocity estimated from OH 1612\,MHz and SiO maser observations is $22 \pm 1$\,km\,s$^{-1}$ \citep{Richards1998MNRAS.299..319R} (all velocities are with respect to the local standard of rest, LSR).

Optical and near-infrared high-resolution imaging and interferometry have revealed a complex distribution of knots and filamentary arcs in the asymmetric reflection nebula around the obscured central star \citep{Kastner1998AJ....115.1592K, Monnier1999ApJ...512..351M, Smith2001AJ....121.1111S}. Images, polarimetric data and radial velocity measurements \citep{Smith2004MNRAS.349L..31S, Humphreys2007AJ....133.2716H, Jones2007AJ....133.2730J} show evidence that the arcs and knots in the ejecta are not only spatially distinct features, but are also kinematically separate from the surrounding diffuse material representing the steady wind \citep{Humphreys2007AJ....133.2716H, Smith2009AJ....137.3558S}. The current understanding is that VY~CMa went through a series of episodic mass ejections with dense cloudlets moving at $\sim$35\,km\,s$^{-1}$ that produced a very high mass-loss rate of $\sim 2\times10^{-3}$\,\Msun\,yr$^{-1}$ between a few hundred and 1\,000 year ago. The diffuse halo might expand at a slightly lower speed ($\sim$25\,km\,s$^{-1}$) and corresponds to the current average mass-loss rate of $\sim3\times10^{-4}$\,\Msun\,yr$^{-1}$ \citep{Smith2009AJ....137.3558S}. Optical and infrared data also provide some evidence for a disk-like geometry near the central object \citep{Herbig1970ApJ...162..557H, Herbig1972ApJ...172..375H, McCarthy1979hars.proc...18M, Efstathiou1990MNRAS.245..275E}. The derived geometry is not always consistent between these studies. $K$-band interferometric measurements indicate a dusty disk oriented east-west \citep{Monnier2004ApJ...605..436M}, but infrared images have been interpreted as evidence for a northeast-southwest axis of symmetry that represents the polar/rotation axis of the star and disk, which is tilted by 15\deg\ to 30\deg\ to our line of sight \citep[for an overview, see][]{Smith2001AJ....121.1111S, Humphreys2007AJ....133.2716H}. Maser data have been interpreted as an expanding outflow, which at sub-arcsecond scale is more extended in the east-northeast and west-southwest direction \citep{Morris1980AJ.....85..724M, Bowers1983ApJ...274..733B, Richards1998MNRAS.299..319R}.
The idea of a preferred northeast-soutwest axis is roughly supported by the Hubble Space Telescope (HST) data of \citet{Smith2001AJ....121.1111S} of which the asymmetric emission in the South-West could be interpreted as being reflection along the lower density bipolar axis oriented northeast-southwest with the south-western lobe closer to us \citep[see also the cartoon of the proposed geometry in Fig.~13 in][]{Smith2009AJ....137.3558S}. The apparent random orientation of the arcs, filaments and knots in the wind of VY~CMa suggest that they were produced by localized mass ejections,  not necessarily aligned with either the star's presumed northeast-southwest axis or its equator, and hence independent of any proposed bipolar geometry.

While data taken at optical and infrared continuum wavelength bands are prone to high extinction effects complicating an unambiguous interpretation of the inner wind structure of VY~CMa, the scattering and extinction in the far-infrared and submillimeter domain are almost negligible. 
The retrieval of the morphological, kinematical and chemical structure of the inner wind region from prior observations was often hampered by a lack of spatial resolution to resolve the different kinematical components. Another difficulty was the association of molecular line emission to specific morphological features which was only based on a tentative interpretation of spatially unresolved velocity features in the broad asymmetric line profiles \citep{Ziurys2007Natur.447.1094Z, Milam2007ApJ...668L.131M, Adande2013ApJ...778...22A}. Modern interferometers offer an angular resolution comparable to the angular scale of some arcs and knots in the CSE of VY~CMa and interferometric far-infrared and submillimeter observations of molecular lines hence provide us with crucial diagnostics to unravel the physical and chemical processes shaping the inner envelope of VY~CMa. With the aim to investigate the spatio-kinematical structure of the nebula around VY~CMa, \citet{Kaminski2013ApJS..209...38K} have used the Submillimeter Array (SMA) to undertake an interferometric spectral imaging survey between 279 and 355\,GHz at a spatial resolution of 0.9\arcsec. Twenty-one molecules were detected, of which 11 show a `point-like' morphology, 4 were indicated as `double' with a second source located to the southwest of the star (at 0\farcs8 to 1\farcs2), one molecule (CN) has a `ring-like' distribution, and 5 showed an `extended' distribution with multiple components. One of the detected molecules is NaCl, showing a `point-like' angular distribution in the SMA data with the exception of the most intense NaCl line which shows weak emission in the southwest clump. 

VY~CMa was observed with ALMA during Science Verification in band~7 and band~9 \citep{Richards2014A&A...572L...9R}. Four NaCl thermal emission lines are detected in the band~7 data, and are spatially resolved at an angular resolution of $\sim$~$0\farcs24\times0\farcs13$ (see Sect.~\ref{Sec:observations}). The benefits of studying the spatio-kinematical and chemical structure of the nebula around VY~CMa via NaCl emission are manifold: (1)\,NaCl is a diatomic molecule with a simple Grotrian-diagram reducing the complexity of radiative transfer analysis and thus helping to retrieve the morphology of the nebula, (2)\,NaCl has a high dipole moment \citep[$\mu=9$\,D,][]{Muller2005JMoSt.742..215M} enabling efficient radiative excitation and hence providing a link between the line emission and the radiation field throughout the nebula, (3)\,while the abundance of NaCl in the wind of VY~CMa is estimated to be some 5 orders of magnitude lower than that of CO \citep{Milam2007ApJ...668L.131M}, its high dipole moment facilitates its detection, and (4)\,as most metal-bearing molecules NaCl is very refractory and is expected to  quite easily form solid-state condensates so that one can investigate the gas-grain chemistry in the envelope. Understanding the gas-grain chemistry in this kind of oxygen-rich winds has a very broad application, since the abundances of the most chemically active species are quite close to the Galactic cosmic abundances and hence provides insight in the chemical processes at work on broad galactic scales.

In Sect.~\ref{Sec:observations} we describe the ALMA observations and some details of the data reduction. Sect.~\ref{Sec:results} provides a detailed description of the spectral line profiles, the spatial distribution of the NaCl emission  seen in the channel maps and a non-exhaustive comparison to other tracers of the morphological structure not yet mentioned in the introduction. In Sect.~\ref{Sec:discussion} we discuss the NaCl excitation conditions, construct a simplified chemical network aiming at unravelling the cycling between gaseous and solid NaCl (Sect.~\ref{Sec:condensation}), and finally embark on an in-depth discussion on the morphology of the inner envelope based on the ALMA NaCl channel maps (Sect.~\ref{Sec:Morphology}). We summarize our conclusions in Sect.~\ref{Sec:conclusions}.

\section{Observations}\label{Sec:observations}

VY~CMa was observed for ALMA Science Verification on 2013 16--19 August using 16--20 12-m antennas on baselines from 0.014--2.7\,km.
The main objective was to map the H$_2$O maser lines at 321 and 325\,GHz (Band~7) and 658\,GHz (Band~9), but several thermal lines identified with various rotational transitions of NaCl, TiO$_2$, SO$_2$, and SiO were also present in the spectral setting in addition to the continuum data. \citet{Richards2014A&A...572L...9R} summarises the observations, data reduction and H$_2$O maser results, while \citet{OGorman2015A&A...573L...1O} describes the dust-dominated continuum in more detail. The TiO$_2$ thermal emission lines are presented by \citet{DeBeck2015}. Two prominent components 
are visible in the continuum dust emission around 321\,GHz. The brightest dust component `C' has a peak position at $\alpha$\,=\,07h22m58.34293s and $\delta$\,=\,$-$25\deg46\arcmin03\farcs2227, while the position of the peak of the second continuum component has been shown to coincide with the location of the stellar position `VY' at $\alpha$\,=\,07h22m58.326s and $\delta$\,=\,$-$25\deg46\arcmin03\farcs042 \citep{Richards2014A&A...572L...9R, OGorman2015A&A...573L...1O} (see Fig.~\ref{Fig:NaCl_312}, Fig.~\ref{Fig:NaCl_dust_maser}). 

Four NaCl lines in the ground or first vibrational state are covered in the ALMA band~7 data\footnote{The identification of the NaCl lines is based on the Cologne Database for Molecular Spectroscopy \citep[CDMS;][]{Muller2001A&A...370L..49M, Muller2005JMoSt.742..215M}.}. All of them are detected, albeit only two are unblended (see Table~\ref{Table:iden_NaCl}). The NaCl v=1 J=25-24 is slightly blended with the TiO$_2$($33_{8,26}-33_{7,27}$) line at 322.613\,GHz \citep{DeBeck2015}, while the NaCl v=0 J=25-24 lines is blended with the 325.153\,GHz H$_2$O maser line \citep{Richards2014A&A...572L...9R} and the SiS v=1 J=18-17 line at 325.059\,GHz (see Sect.~\ref{Sec:spectral_profiles}). No rotational lines in the ground or first vibrational state of the minor isotopologue Na$^{37}$Cl were covered in the current ALMA spectral setup in band~7. The NaCl v=3 J=51-50 line at 657.134\,GHz was also detected, but is severely blended with the much stronger H$_2$O maser line at 658\,GHz, making a quantitative analysis impossible and hence it is not further discussed.

The synthesized beamsize for the band~7 data is $\sim 0\farcs24 \times 0\farcs13$\ (see Table~\ref{Table:iden_NaCl}). The stellar continuum was detectable in the individual channels at all frequencies, so that the relative alignment of all lines is good to $\sim$2\,mas \citep{Richards2014A&A...572L...9R}. The continuum was subtracted before making the final line cubes. All data were adjusted to constant velocity with respect to the Local Standard of Rest.

Image cubes were made at 1\,km s$^{-1}$ spectral resolution
around the NaCl frequencies listed in Table~\ref{Table:iden_NaCl}.  The higher
$\sigma_{\mathrm{rms}}$ noise around 325\,GHz is due to the poor
atmospheric transmission in this region.  The noise in the brightest
channels is slightly higher in all cases due to dynamic range
limitations.  The shortest baseline was 14\,m and inspection of the
continuum visibility amplitudes against baseline length shows that the
flux density remains quite steady out to 170\,m at 321--325\,GHz,
suggesting that we recover all the flux on scales $<13$\arcsec, with
reliable imaging on scales up to at least 8\arcsec.  The absolute
astrometric accuracy of the lines described here is 35 mas and the
flux density scale is accurate to at least 10\%.  
\section{Results}\label{Sec:results}

\subsection{Spectral line profiles}\label{Sec:spectral_profiles}

\begin{table*}[htp]
 \caption{NaCl lines in the spectral windows covered by the ALMA observations. Listed are the transition, rest frequency, lower state energy, Einstein A-values from CDMS, the rms noise, the synthesized beam, and comments concerning line blending.}
 \label{Table:iden_NaCl}
 \begin{tabular}{ccccccl}
 \hline \hline
  Transition	& $\nu_{\rm{lab}}$	& $E_{\rm{low}}$ 	& A 			& Rms noise 		& Beam 	& Comment\\
		& [GHz]			& [cm$^{-1}$] 			& sec$^{-1}$ 	&[mJy~beam$^{-1}$ km\,s$^{-1}$]	& [\arcsec$\times$\arcsec, P.A.] 	&\\
 \hline
 v=1 J=24-23 & 309.787		& 480.09 				& 0.017		& 19			& 0.232$\times$0.131, 28.27\deg 	& unblended \\
 v=0 J=24-23 & 312.110		& 119.83 				& 0.013		& 30			& 0.239$\times$0.134, 29.48\deg 	& unblended \\
 v=1 J=25-24 & 322.650		& 490.42				& 0.019		& 27			&0.222$\times$0.126, 27.55\deg	& red wing slightly blended by TiO$_2$ line\\
 v=0 J=25-24 & 325.069		& 130.24				& 0.015		& 195			&0.217$\times$0.119, 29.27\deg	& blue wing blended by H$_2$O maser\\
	    &			&			&					&				&  & + blend with SiS v=1 J=18-17 \\
 \hline
 \end{tabular}
\end{table*}

\begin{figure}[htp]
\centering\includegraphics[width=0.35\textwidth,angle=90]{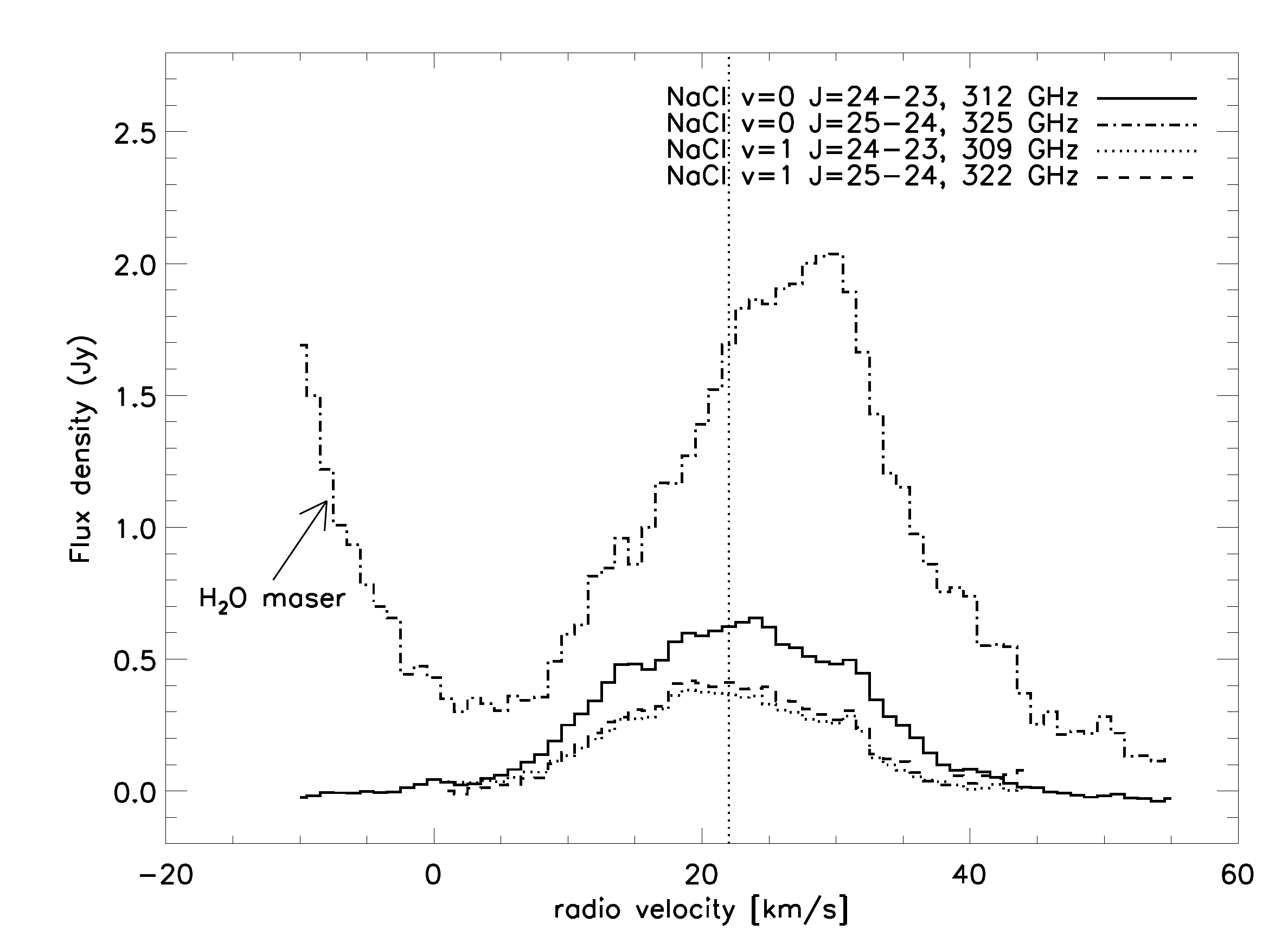} 
\caption{Flux densities for the four NaCl lines detected in the ALMA band~7 data. Flux densities are for an aperture with circular diameter of 1\arcsec\ centered at `VY'. The LSR velocity of 22\,km\,s$^{-1}$ is indicated by the vertical dotted line. Line blending causes the flux density for the 325\,GHz NaCl v=0 J=25-24 line to be skewed and to be too high with an expected contribution of the SiS v=1 J=18-17 line of $\sim$1.1\,Jy (see text for more details).}
\label{Fig:spectra_all_NaCl}
\end{figure}

Fig.~\ref{Fig:spectra_all_NaCl} shows the flux densities extracted for a 1\arcsec\ diameter region around the stellar position `VY' for the four NaCl lines; no NaCl v=1 emission is detected beyond this aperture. The four NaCl lines are centered around the $v_{\rm{LSR}}$ of $\sim$22\,km\,s$^{-1}$  \citep{Richards1998MNRAS.299..319R, Tenenbaum2010ApJS..190..348T, Kaminski2013ApJS..209...38K, Richards2014A&A...572L...9R} and show similar line shapes ranging from $\sim$6\,km\,s$^{-1}$ to $\sim$40\,km\,s$^{-1}$. Given the similar energy levels, quantum numbers, and Einstein-A coefficients, one would expect similar line intensities and spatial distributions for both lines in the same vibrational level. While this holds for the NaCl lines in the first vibrational level, the 325\,GHz NaCl J=25-24 line in the vibrational ground state is $\sim$1.4\,Jy higher than the 312\,GHz NaCl v=0 J=24-23 peak. This appears to be due to blending.  The 325.153\,GHz water maser (with a $\sim$500\,Jy peak) has a very broad profile, with a red-shifted wing $>$70\,km s$^{-1}$ and contributes roughly 0.2\,Jy to the low-velocity wing of the 325.069\,GHz NaCl line. The 325.059\,GHz SiS v=1 J=18-17 line is even closer to the NaCl line (peaks separated by 9.2\,km s$^{-1}$). To estimate the line strength of the SiS v=1 J=18-17, we use the results of \citet{Kaminski2013ApJS..209...38K} who made a spectral line survey of VY~CMa between 279 and 355\,GHz using the Submillimeter Array (SMA).  They were unable to detect the  325.059 GHz\,SiS line due to the poor atmospheric transmission but, using a 1\arcsec\ radius aperture, they measured maximum flux densities of 0.96\,Jy for the SiS v=1 J=16-15 line and 0.97\,Jy for the J=17-16 line. The estimated maximum flux for the SiS v=1 J=19-18 is indicated to be uncertain, with a value around 1.38\,Jy. From these results, we estimate that the SiS v=1 J=18-17 line can contribute $\sim$1.1\,Jy to the flux density profile of the 325\,GHz NaCl line, mostly on the higher-velocity side. This explains why this line is skewed to the red-shifted side of the stellar velocity.  If  1.1--1.3\,Jy is subtracted from the apparent 325\,GHz NaCl line profile, the remaining peak would be close to the 0.6\,Jy peak of the 312\,GHz NaCl line. After correction for line blending, both v=1 lines have flux densities within a factor of $\sim$2 of the v=0 lines despite an energy separation of $\sim$360\,cm$^{-1}$, which is caused by the dominantly radiative excitation mechanism.

\subsection{Spatial distribution}\label{Sec:spatial_distribution}

\begin{figure*}[htp]
\begin{center}
\includegraphics[width=0.8\textwidth]{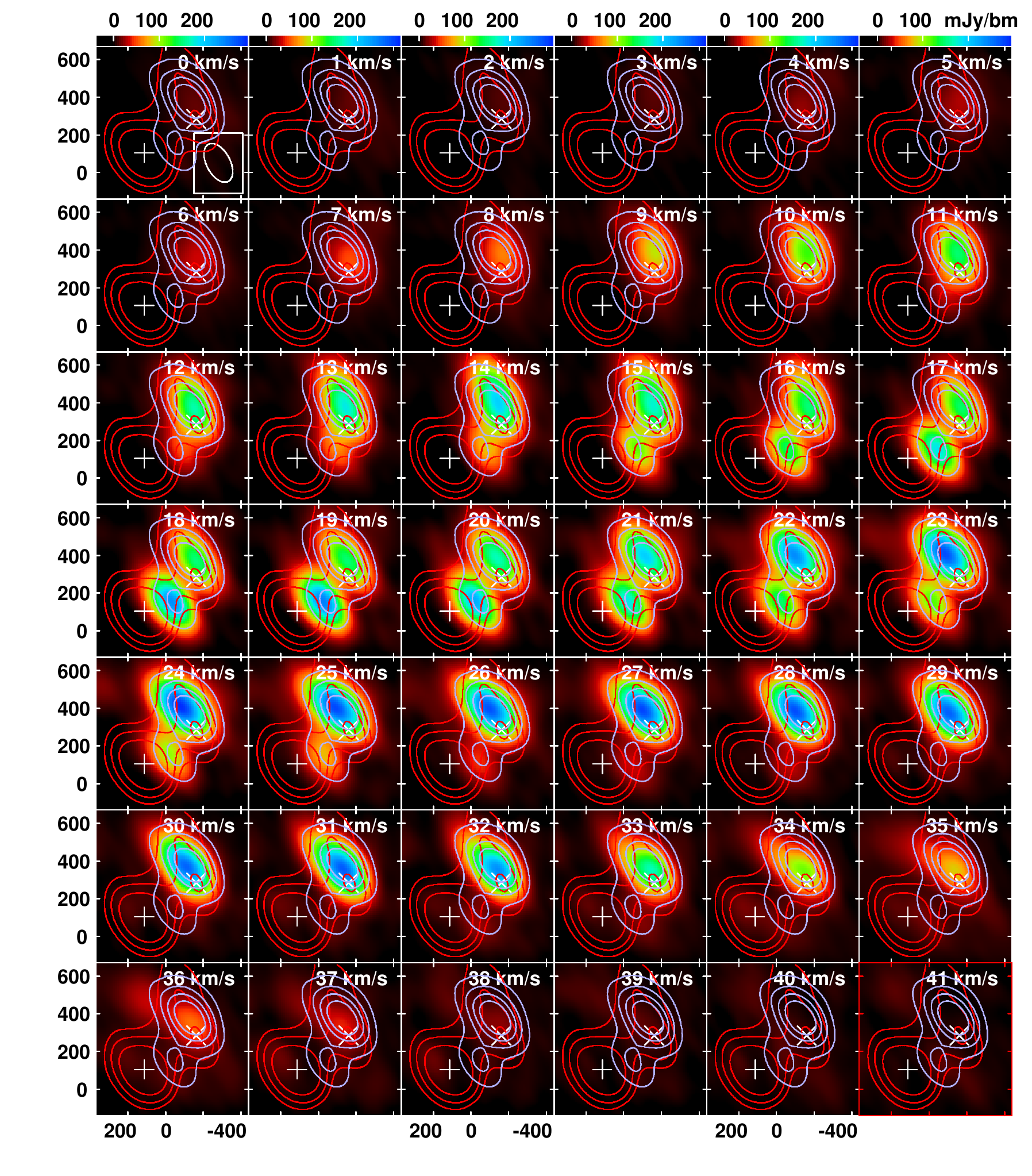}
\end{center}
\caption{Channel maps (colour scale) of the NaCl v=0 J=24-23 emission at 312\,GHz at a 1\,km s$^{-1}$ velocity resolution for the central 0\farcs8$\times$0\farcs8 field around `VY' and `C'. Grey contours are the integrated NaCl line strength at $(1, 2, 3, 4)\times 1$\,Jy/beam km\,s$^{-1}$, red contours are the dust continuum measured with ALMA at 321\,GHz at $(1, 2, 3, 4) \times 18$\,mJy/beam \citep{Richards2014A&A...572L...9R, OGorman2015A&A...573L...1O}. The stellar position `VY' is indicated with a white cross, and the peak of the continuum emission `C' is indicated with a white plus-sign. The ordinate and co-ordinate axis give the offset of the right ascension and declination, respectively, in units of milli-arcseconds.}
\label{Fig:NaCl_312}
\end{figure*}

\begin{figure*}[htp]
\sidecaption
\includegraphics[width=120mm]{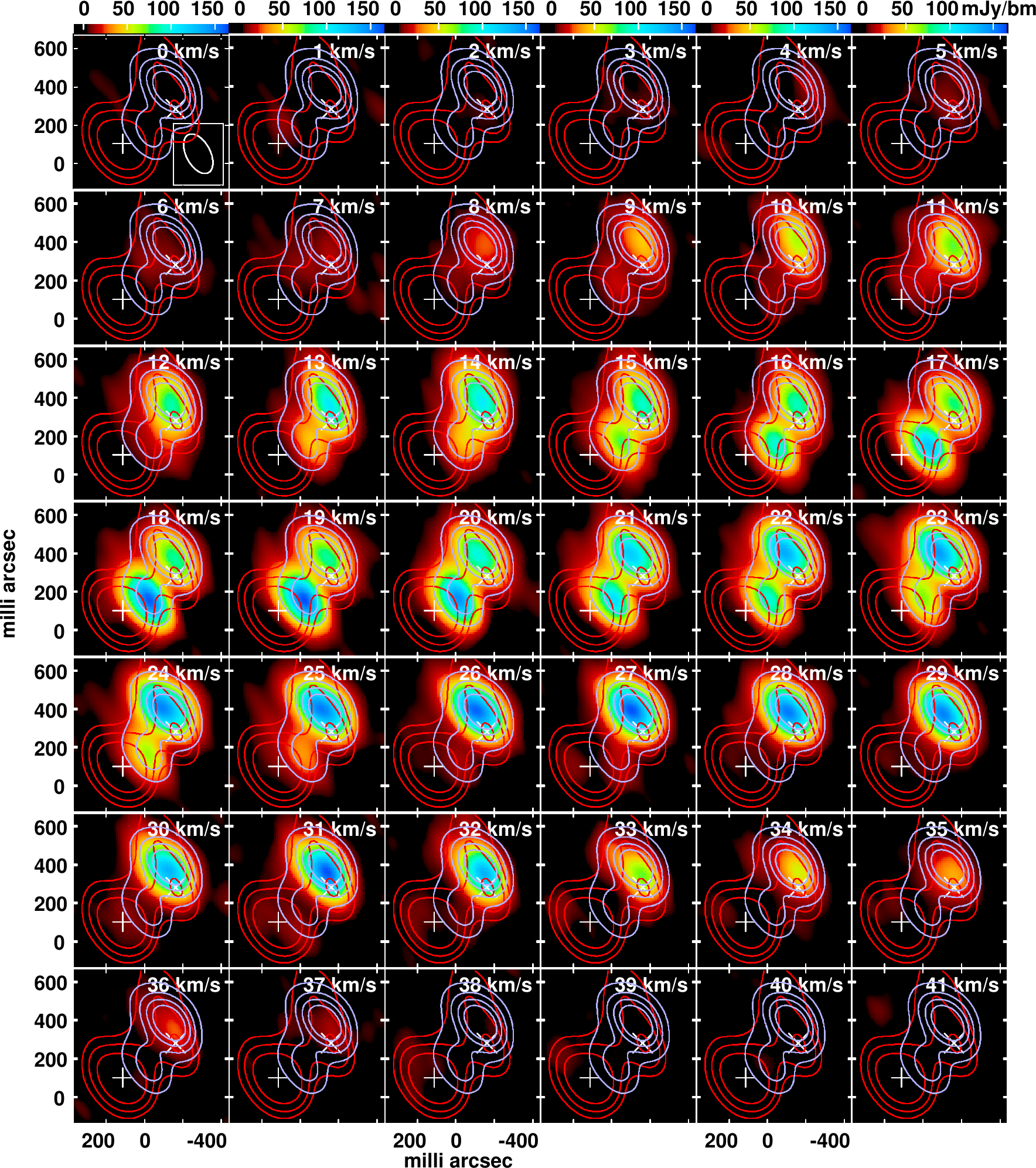}
\caption{Same as Fig.~\ref{Fig:NaCl_312} but for the NaCl v=1 J=24-23 line at 309\,GHz.}
\label{Fig:NaCl_309}
\end{figure*}

\begin{figure*}[htp]
\sidecaption
\includegraphics[width=120mm]{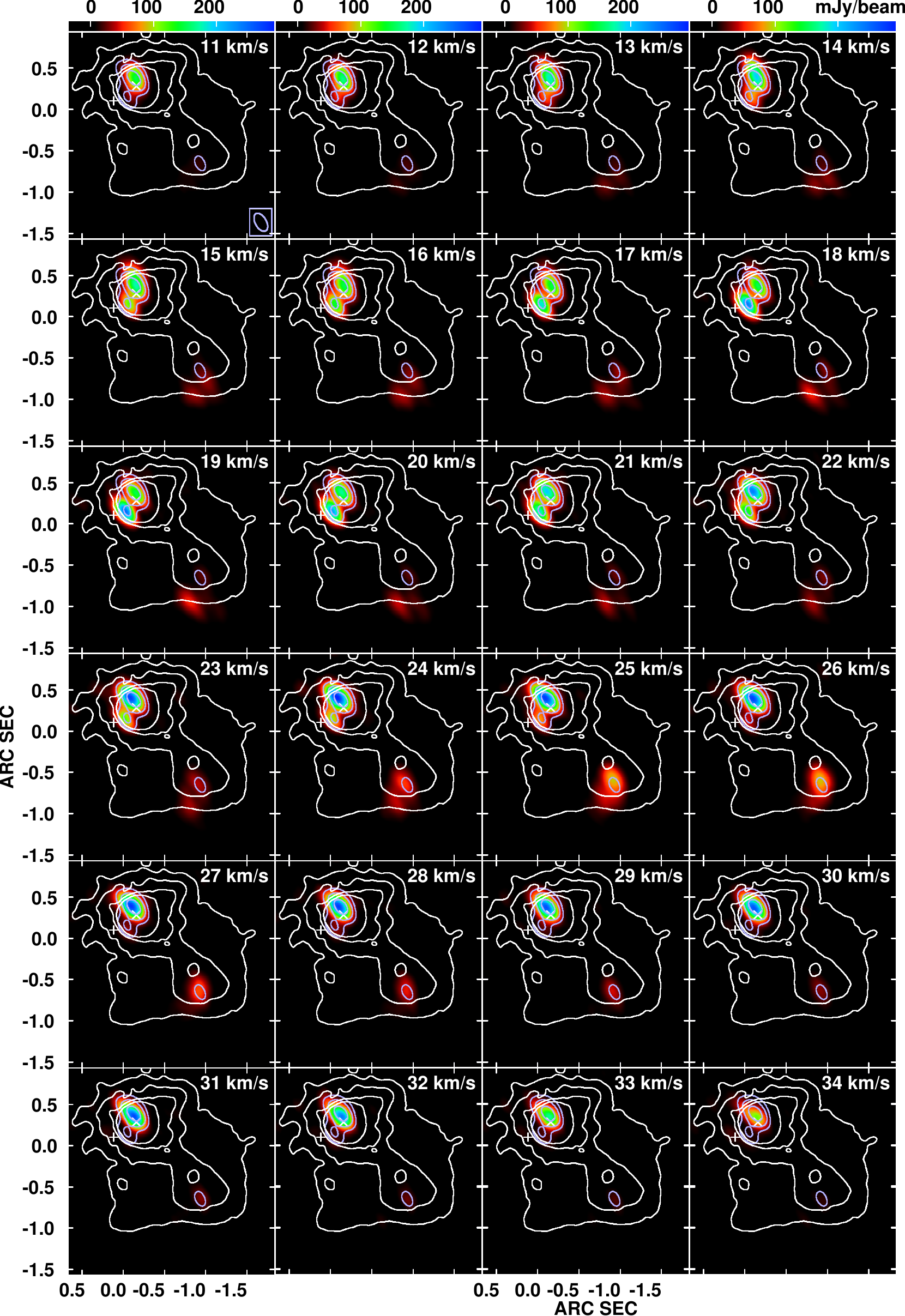}
\caption{Same as Fig.~\ref{Fig:NaCl_312}, but showing the central 2\farcs5$\times$2\farcs5 region. The white contours show the HST/WFPC2 data in the F1042M filter around 1.018\,$\mu$m at $(1, 2, 4, 8, 16)\times110$ counts \citep{Smith2001AJ....121.1111S}.}
\label{Fig:NaCl_312_HST}
\end{figure*}

\begin{figure}[htp]
\centering \includegraphics[width=0.48\textwidth]{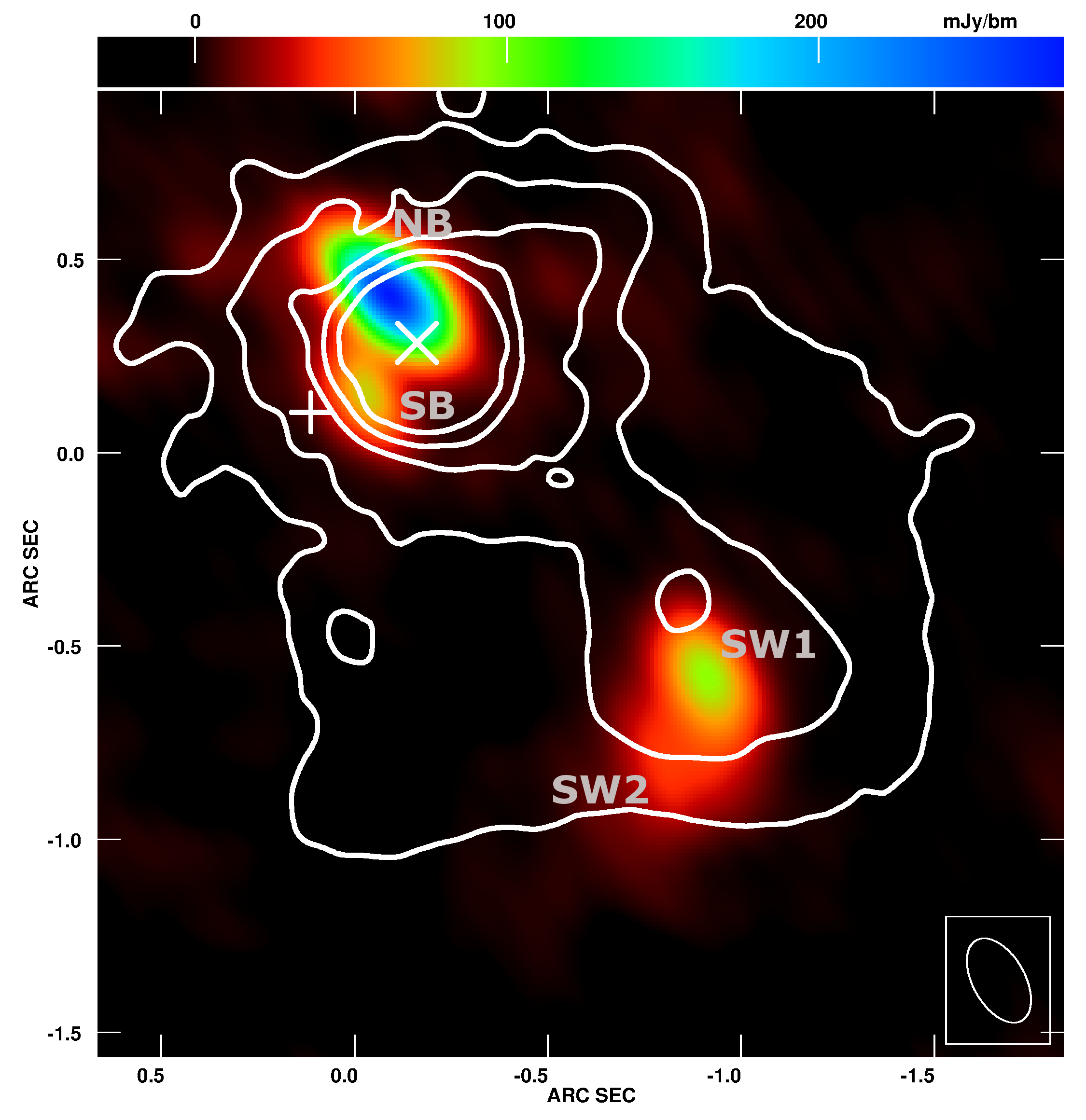}
\caption{ALMA NaCl 312\,GHz channel map at 25\,km\,s$^{-1}$ (colour scale) overlaid with the HST/WFPC2 F1042M data of \citet{Smith2001AJ....121.1111S} (in contours at $(1, 2, 4, 8, 16)\times110$ counts). The cross sign indicates the position of the central star VY CMa, noted `VY' in the text, as derived from the ALMA H$_2$O maser and continuum data \citep{Richards2014A&A...572L...9R, OGorman2015A&A...573L...1O}, the plus-sign indicates the maximum of the ALMA dust continuum data, noted as `C' in the text, \citep{OGorman2015A&A...573L...1O}. The nomenclature of the different spatial regions refers to the northern clump 'NB', the southern clump 'SB', the brightest part of the south-western extension 'SW1', and the faintest part of the north-western extension 'SW2'. The ordinate and co-ordinate axis give the offset of the right ascension and declination, respectively, in units of arcseconds.}
\label{Fig:NaCl_312_25km_HST}
\end{figure}

\begin{figure}[htp]
\centering\includegraphics[width=0.35\textwidth,angle=90]{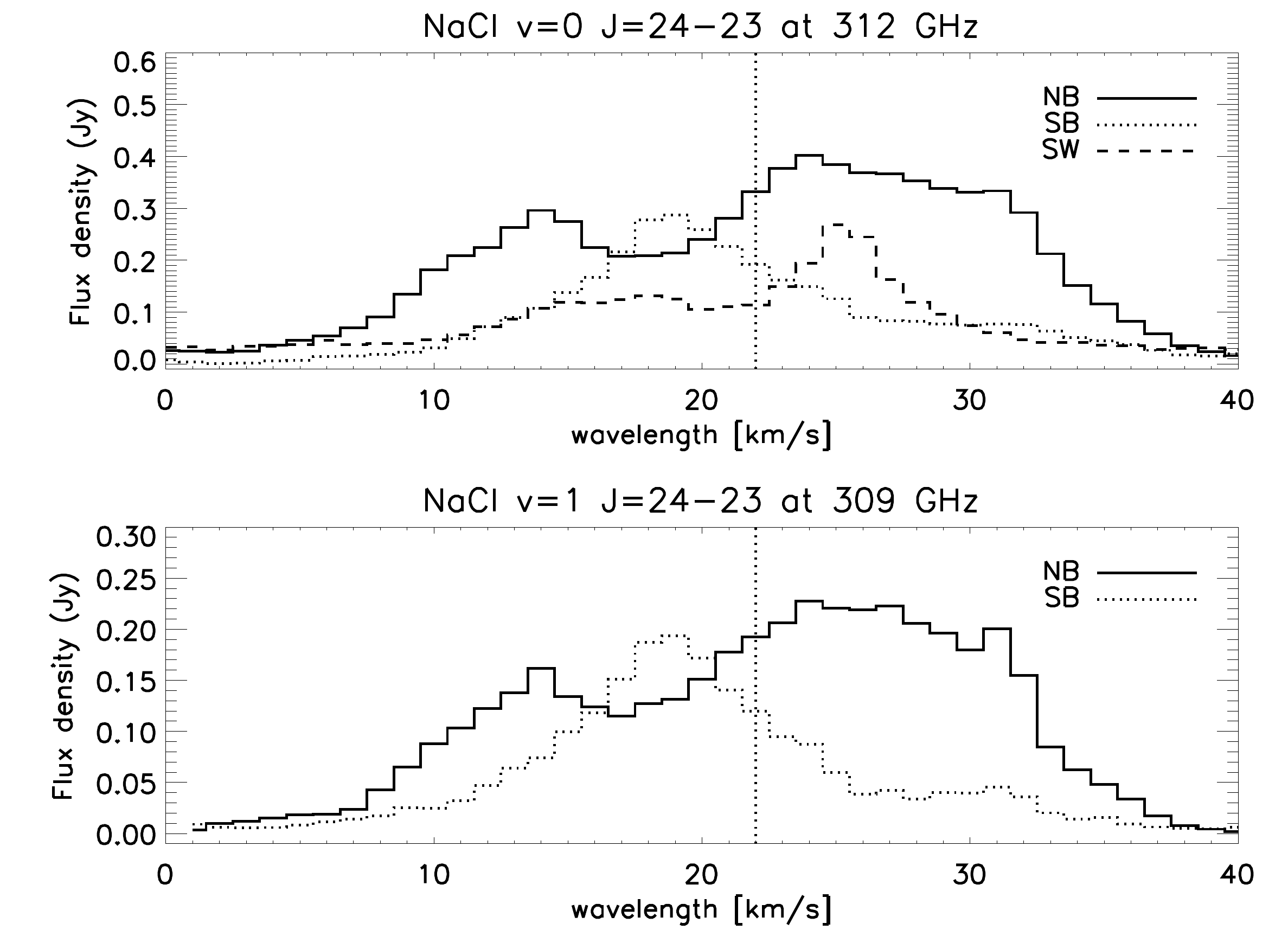} 
\caption{Flux densities for the NaCl J=24-23 lines in the ground-vibrational state v=0 at 312\,GHz (upper panel) and in the first vibrational state v=1 at 309\,GHz (lower panel). Flux densities are extracted in the north clump (NB, full line) at (07h22m58.326s,$-$25\deg46\arcmin02\farcs9) for an aperture of (0\farcs394$\times$0\farcs373), in the south clump (SB, dotted line) at (07h22m58.332s,$-$25\deg46\arcmin03\farcs118) for an aperture of (0\farcs36$\times$0\farcs36), and in case of the 312\,GHz line in the south-west clump (SW, dashed line) at (07h22m58.273s,$-$25\deg46\arcmin04\farcs028) for an aperture of (0\farcs57$\times$0\farcs73). The LSR velocity of 22\,km\,s$^{-1}$ is indicated by the vertical dotted line.}
\label{Fig:spectra_NaCl}
\end{figure}

We created moment-zero maps (total intensity) for each of the NaCl lines. The NaCl lines in a given vibrational level show similar spatial distributions. Since the NaCl J=25-24 lines are blended in both vibrational levels, we discuss the spatial distribution of NaCl on the basis of the J=24-23 rotational lines; the channel maps for the NaCl J=25-24 lines are shown in Appendix~\ref{App:extra_channel_maps}. A zoom for the inner 0.8\arcsec\ around the stellar position `VY' is shown in Fig.~\ref{Fig:NaCl_312} and Fig.~\ref{Fig:NaCl_309}. In this inner 0.8\arcsec\ region, the spatial distribution of both lines show a similar bimodal distribution with two clumps at small offsets from the stellar position. For most of the channel maps, the northern clump (`NB' from here onward) is brighter than the southern clump (`SB' from here onward); see also Fig.~\ref{Fig:NaCl_312_25km_HST} for an indication of the nomenclature used in this paper. In addition, both rotational NaCl lines in the ground-vibrational state display a third emission component located at $\sim$1.2\arcsec\ to the South-West (`SW' clump); see Fig.~\ref{Fig:NaCl_312_HST}.

For the NaCl lines in the first vibrational state, all the emission is concentrated in `NB' and `SB'; the other two lines in the ground vibrational state show two (one) `SW' component(s) at 312 (325)\,GHz. The two `SW' components for the NaCl v=0 J=24-23 line at 312\,GHz are called `SW1' and `SW2' (see Fig.~\ref{Fig:NaCl_312_25km_HST}), with `SW1' being brighter than `SW2'. We measured the sizes and positions of each spatial region by fitting two  dimensional elliptical Gaussian components (see Table~\ref{Table:spatial}).  The position uncertainties given are estimated conservatively from the noise as (mean beam size)/(signal to noise ratio) but do not include the 2\,mas line-to-line alignment uncertainty (the 35\,mas absolute astrometric uncertainty does not affect alignment of these NaCl lines). The component sizes are the FWHM (full width at half maximum) deconvolved from the beam where possible; only `NB' is consistently fully resolved and the errors given take into account that the sizes of the other components are in some cases upper limits.

\begin{table*}[htp]
\caption{Information of the different spatial components seen in the ALMA NaCl channel maps. The first column indicates which spatial component was fitted with a 2-dimensional (2D) Gaussian function, the second column gives the peak of the fitted function, the third column the total flux, the fourth, fifth, and sixth column the right ascension, declination, and positional uncertainty of the center of the elliptical Gaussian fit, the seventh, eight and ninth column the major and minor axis of the deconvolved 2D Gaussian fit with the uncertainty, and the last two columns list the position angle and its uncertainty of the major axis.}
\label{Table:spatial}
\setlength{\tabcolsep}{1.mm}
\begin{tabular}{lcrrrrrrrrr}
\hline \hline
Region$^{(a)}$ &   \multicolumn{1}{c}{Peak}  &   \multicolumn{1}{c}{Tot flux} & \multicolumn{1}{c}{RA} &   \multicolumn{1}{c}{Dec}  &    \multicolumn{1}{c}{Pos err} &  \multicolumn{1}{c}{Maj ax} & \multicolumn{1}{c}{Min ax} & \multicolumn{1}{c}{Size err} & \multicolumn{1}{c}{PA}  & \multicolumn{1}{c}{PA err}\\
     & \multicolumn{1}{c}{[Jy/beam km\,s$^{-1}$]}  & \multicolumn{1}{c}{[Jy km\,s$^{-1}$]}   &  \multicolumn{1}{c}{[h:m:s]}     & \multicolumn{1}{c}{[\deg:\ \arcmin:\ \arcsec]}    & \multicolumn{1}{c}{[mas]}   &  \multicolumn{1}{c}{[mas]} &  \multicolumn{1}{c}{[mas]} & \multicolumn{1}{c}{[mas]}  & \multicolumn{1}{c}{[degrees]} & \multicolumn{1}{c}{[degrees]} \\
     \hline
\multicolumn{11}{l}{NaCl $v=1 J=24-23$ at 309\,GHz}\\    
	NB  &  2.831 &   5.162 &  07:22:58.3253 &  $-$25:46:02.937 & 2 &  189 &    121 &  5  &  59 &   4 \\
 	SB  &  1.248  &  2.144 &  07:22:58.3340 &  $-$25:46:03.162 & 4 &  196 &    105 & 10   & 48  & 10 \\
\hline
\multicolumn{11}{l}{NaCl $v=0 J=24-23$ at 312\,GHz}\\  
     NB  &  5.327  & 10.539  & 07:22:58.3253  & $-$25:46:02.942 & 2  & 213    & 141  & 35  & 50 & 45 \\
     SB* &  2.134  &  2.607  & 07:22:58.3311  & $-$25:46:03.195 & 4  & 138    & 134  & 70  & 63 & 90 \\
     SW1* &  1.184  &  3.052  & 07:22:58.2663  & $-$25:46:03.992 & 7  & 281    & 174 & 160  &  9 & 160 \\
     SW2* &  0.856  &  1.860  & 07:22:58.2779  & $-$25:46:04.253 & 9  & 276    & 131 & 200  & 42 & 180 \\
\hline
\multicolumn{11}{l}{NaCl $v=1 J=25-24$ at 322\,GHz} \\       
     NB   & 2.883   & 6.295  & 07:22:58.3259  & $-$25:46:02.946 & 3  & 181    & 166  & 7    & 85 & 20 \\
     SB*  & 1.377   & 1.855  & 07:22:58.3338  & $-$25:46:03.181 & 5  & 167    & 126 & 13   & 57 & 30 \\
\hline
\multicolumn{11}{l}{NaCl $v=0 J=25-24$ at 325\,GHz$^{(b)}$}\\       
     NB  & 13.924 &  55.232 &  07:22:58.3272  & $-$25:46:03.009  & 3  & 328    & 213 & 200 & 156 & 120 \\
     SB*  & 1.621  &  1.621 &  07:22:58.3300  & $-$25:46:03.230 & 26  & 217    & 119 & 200  & 61 & 180 \\
    SW  & 0.588  &  1.843 &  07:22:58.2712  & $-$25:46:04.271 & 72  & 292    & 150 & 200 & 144 & 180 \\
\hline
\end{tabular}
\tablefoot{
\tablefoottext{a}{Indication of the different spatial regions (see Fig.~\ref{Fig:NaCl_312_HST}). A `*' denotes that the region is only partly resolved; i.e.\ the deconvolved size is an upper limit.}
\tablefoottext{b}{Strongly blended with H$_2$O and SiS line.}
}
\end{table*}

The brightest `NB' component is aligned to within 8\,mas in both coordinates at 309, 312 and 322\,GHz, which is within the 3\,sigma  position uncertainties. The fainter `SB' component is aligned to within 40\,mas; although this could be significant it is more probably due to fitting uncertainties in the case of a non-Gaussian flux distribution.  The average separations of `NB' and `SB' from the stellar position, `VY', are  100\,mas (or $\sim$18\,\Rstar) and 190\,mas (or $\sim$34\,\Rstar), respectively.  The separation between `NB' and `SB' is 265\,mas ($\sim$48\,\Rstar). The relative uncertainty in each case is 10\,mas ($\sim$2\,\Rstar).  The brighter `SW' component, `SW1', seen at 312\,GHz is 1.22\arcsec\ ($\sim$220\,\Rstar) from VY; the fainter `SW2' component is 0.3\arcsec\ further away.   Formally, the angular separations are a lower limit to the physical separation, but since the  NaCl clumps' velocities span the stellar velocity, these are probably close to the actual separations.

`NB' and `SB' and a single `SW' clump are seen in the NaCl v=0 J=25-24 line at 325\,GHz.  The uncertainties listed in Table~\ref{Table:spatial} are a few times greater, due both to the noise level which is 6-10 times higher and blending with the other lines, in particular with the water maser which is widely distributed around  the`NB' and `SB' clumps and beyond but not towards the `SW' clump.

The northern clump is visible in the channel maps for all observed NaCl transitions with velocities between 6 and 37\,km\,s$^{-1}$. The flux density in `NB' shows two local maxima, one around 14\,km\,s$^{-1}$ (`NB$_{\rm{blue}}$') and the other around 24\,km\,s$^{-1}$ (`NB$_{\rm{red}}$'); see Fig.~\ref{Fig:spectra_NaCl}.  Emission from `SB' arises in channel maps with velocities between 12 and 26\,km\,s$^{-1}$. The flux density of `SB' reaches a maximum around 19\,km\,s$^{-1}$, i.e.\ coinciding with a local minimum in the flux density of `NB'. The 'SW' extension in the ground vibrational lines is visible for velocity channels between 12-33\,km\,s$^{-1}$ (see Fig.~\ref{Fig:NaCl_312_HST}) with a maximum flux density around 25\,km\,s$^{-1}$ (see Fig.~\ref{Fig:spectra_NaCl}).

\subsection{Morphological comparison to other tracers of the envelope} \label{Sect:comp_others}

In this section we compare the NaCl emission seen in the ALMA channel maps to some other 
results presented in the literature. We only focus on observational results already retrieved 
from ALMA data and on studies which discuss the NaCl emission and/or the extension seen at 
1.2\arcsec\ to the south-west.

The maximum known angular extent of the molecular envelope around VY~CMa as traced by 
CO \citep{Muller2007ApJ...656.1109M, Kaminski2013ApJS..209...38K} and OH maser emission 
\citep{Richards1997PhDT........20R} is $\sim$10\arcsec\ and the typical velocity span is 
from $-$15 to $+60$\,km\,s$^{-1}$, with several species observed closer to the central star 
showing a more extended red wing up to $+105$\,km\,s$^{-1}$ \citep[see, e.g.][]{Kaminski2013ApJS..209...38K, DeBeck2015}. 
However, where imaging is available, these extreme velocities appear at only a few position 
angles, and most of the molecular shell appears to show radial acceleration out to a terminal 
velocity of 35--45\,km\,s$^{-1}$ \citep{Decin2006A&A...456..549D, Muller2007ApJ...656.1109M, 
Ziurys2007Natur.447.1094Z}\footnote{Note that \citet{Smith2009AJ....137.3558S} mentioned a 
terminal velocity for the diffuse cloud representing the steady wind of 25\,km\,s$^{-1}$; 
see Sec~\ref{Sec:introduction}.}. The NaCl emission detected by ALMA band~7 has hence a much 
more compact distribution and while its emission extends significantly beyond the estimated 
dust formation radius of 35-55\,mas 
\citep{Danchi1994AJ....107.1469D, Monnier1999ApJ...512..351M, Smith2001AJ....121.1111S, 
Decin2006A&A...456..549D} it does not reach the terminal velocity.

Fig.~\ref{Fig:NaCl_312}, Fig.~\ref{Fig:NaCl_309}, Fig.~\ref{Fig:NaCl_322}, and 
Fig.~\ref{Fig:NaCl_325} compare the NaCl emission to the dust emission detected in the 
ALMA band~7 band. The emission of the brightest NaCl components (`NB' and `SB') and the dust 
shows a similar central `waist' along a NE-SW axis, albeit the peak position of the NaCl 
clumps is to the north of 'VY' and NW of 'C' and are clearly offset from the dust continuum peaks.  

\begin{figure}[htp]
\includegraphics[width=0.48\textwidth]{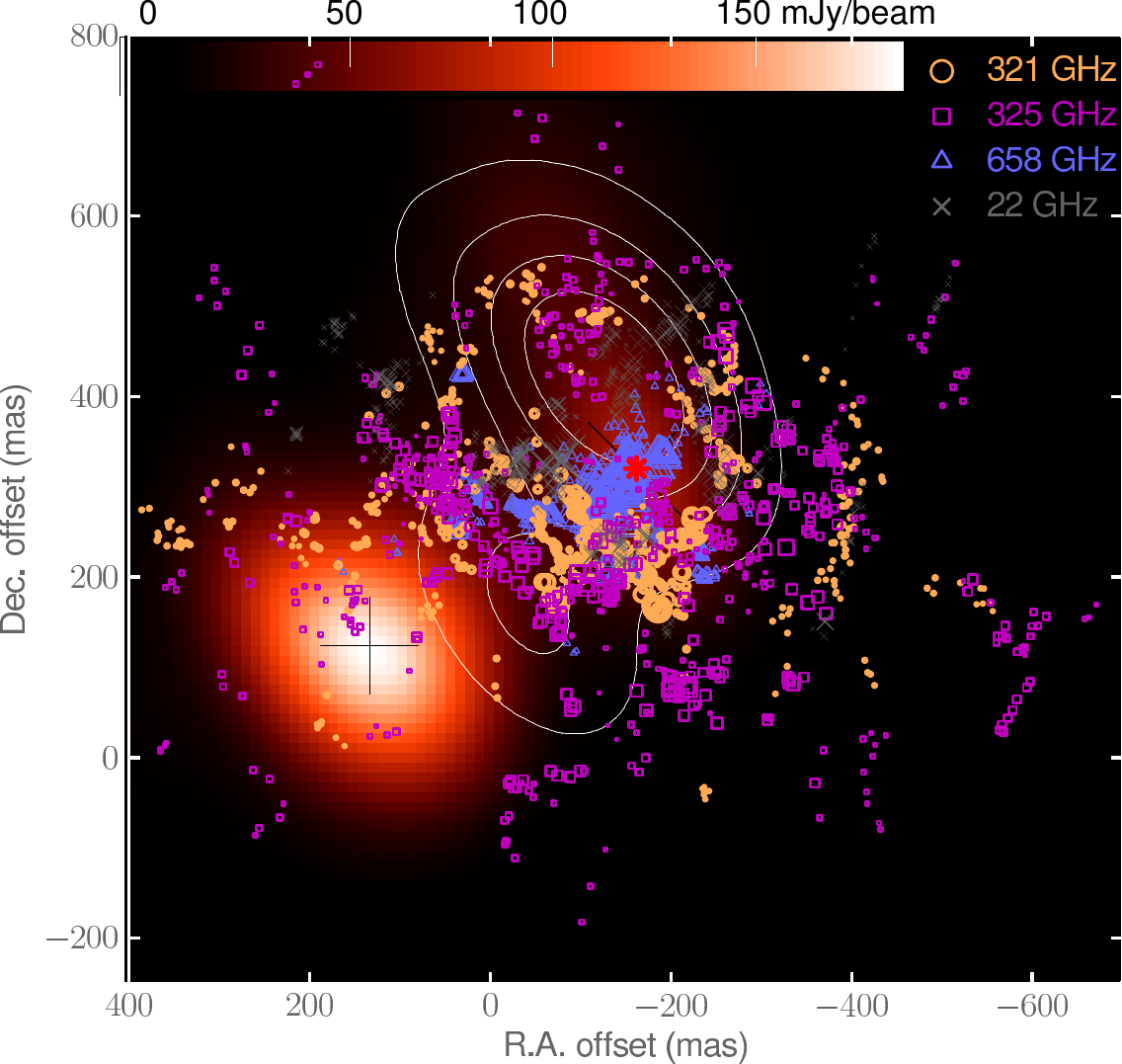}
\caption{The H$_2$O maser emission \citep{Richards1998MNRAS.299..319R, Richards2014A&A...572L...9R} is shown on top of the NaCl 312\,GHz integrated line emission (gray contours, zeroth moment at 0.25, 0.5, 1, 2, 3, 4 Jy km\,s$^{-1}$) and the ALMA dust continuum emission \citep[colour scale;][]{OGorman2015A&A...573L...1O}. }
\label{Fig:NaCl_dust_maser}
\end{figure}

As shown by \citet{Richards2014A&A...572L...9R} the H$_2$O 658, 321, and 325\,GHz maser emission cluster around the stellar position `VY', with another accumulation in the `valley' between `C' and `VY' (see Fig.~\ref{Fig:NaCl_dust_maser}). \citet{Richards1998MNRAS.299..319R} found the 22\,GHz H$_2$O masers to show a generally expanding envelope, with two `ears' outside the main 22\,GHz shell, which tentatively was interpreted as being suggestive for a biconical outflow. The south clump (`SB') of the NaCl emission lies in the `valley' between the two continuum peaks, to the south of some of the most intense water maser emission. The 22\,GHz and 321\,GHz H$_2$O maser features at the greatest angular separations lie approximately east and west or to the southwest.

\citet{Milam2007ApJ...668L.131M} observed several rotational lines in the ground vibrational state of NaCl using single-dish telescopes with beam sizes between 29\arcsec--44\arcsec. The NaCl emission was spatially unresolved and modelling the observed spectra resulted in an abundance relative to H$_2$ of $5 \times 10^{-9}$ with a source size of 0.5\arcsec. \citet{Kaminski2013ApJS..209...38K} observed several rotational lines of NaCl in vibrational states with $v \le 3$ at an angular resolution of $\sim$0.9\arcsec\ using the SMA. The most intense lines were partially resolved, with the emission being slightly elongated with mean deconvolved FWHMs of $(0\farcs42 \pm 0\farcs05) \times (0\farcs30 \pm 0\farcs04)$ and a position angle of $-33 \pm 2$\deg. Weak emission in the SW clump at $\sim$1\arcsec\ is seen only in the most intense line of NaCl and could not be mapped with the SMA. Thanks to the high spatial resolution of ALMA, the NaCl emission is for the first time well resolved  and a clear distinction can be made between emission seen in the bright northern clump, the slightly weaker southern clump, and the weak south-western extension (see Table~\ref{Table:spatial} in Sect.~\ref{Sec:spatial_distribution}), the latter only detected for the NaCl lines in the ground-vibrational state. Based on the NaCl $v=0$ $J=24-23$ transition --- the only unblended line for which the emission traces all the different spatial components --- the northern clump has a deconvolved size of $(0\farcs213 \times 0\farcs141)$, the upper limit for the size of the southern clump is $(0\farcs138 \times 0\farcs134)$, while the two components in the south-western extension have sizes with upper limits being $(0\farcs281 \times 0\farcs174)$ and $(0\farcs276 \times 0\farcs131)$ (see Table~\ref{Table:spatial}).

A similar `SW' extension/clump at $\sim$1\arcsec\ for a position angle (PA) of $\sim$220\deg\ w.r.t.\ `VY' is seen in some ALMA TiO$_2$ channel maps \citep{DeBeck2015} and is also reported by \citet{Kaminski2013ApJS..209...38K} for some sulfur bearing species (CS, NS, SiS, and H$_2$S) and the most intense NaCl (J=26-25) line. The line profile of the SMA H$_2$S and CS emission in the `SW' clump seen is rectangular with FWHM of 15\,km\,s$^{-1}$. The emission of this `SW' component very well matches the south-west extension seen in the Hubble Space Telescope (HST) data at wavelengths between 1--2.14\,$\mu$m that image the reflection nebula around VY~CMa \citep[see Fig.~\ref{Fig:NaCl_312_25km_HST};][]{Smith2001AJ....121.1111S} and with other near-infrared and polarimetric data as for instance presented by \citet{Monnier1999ApJ...512..351M} and \citet{Jones2007AJ....133.2730J}. The direction of this southwest extension/knot is roughly perpendicular to the possible density-enhanced equatorial plane as deduced from the one-sided reflection nebula seen in the visible and the maser elongation (see Sect.~\ref{Sec:introduction}). More recently, \citet{Shenoy2013AJ....146...90S} re-confirmed the existence of a peculiar south-west clump at approximately 1.4\arcsec\ to the southwest of the star using LBT/LMIRCam data. \citet{Humphreys2007AJ....133.2716H} studied the total space motions and directions of several features in the envelope of VY~CMa. They deduced that the southwest clump is slightly redshifted w.r.t.\ the systemic velocity and is moving away from us at an angle of $+8$\deg\ with a total velocity of only $\sim$18\,km\,s$^{-1}$, of which the transverse component is $\sim$17.7\,km\,s$^{-1}$ and with a negligible radial component of $\sim$2.5\,km\,s$^{-1}$. This derived redshifted behaviour is in accord with the ALMA NaCl flux density shown in Fig.~\ref{Fig:spectra_NaCl} which shows a peak around 25\,km\,s$^{-1}$, or 3\,km\,s$^{-1}$ greater than the stellar velocity of 22\,km\,s$^{-1}$. However, this is not in agreement with the ALMA TiO$_2$ emission showing up blue shifted \citep{DeBeck2015} (see Sect.~\ref{Sec:localized}).

\section{Discussion}\label{Sec:discussion}

Metal-bearing molecules are rarely observed in the gas-phase in the ISM due to their large refractory character that makes them to easily form solid condensates.
Metal halides are among the first metal-containing molecules detected in space. The alkali halide salt molecule, NaCl, was already detected in VY~CMa and in the oxygen-rich AGB star IK~Tau \citep{Milam2007ApJ...668L.131M, Kaminski2013ApJS..209...38K}. The high dipole moment of NaCl is favourable for radiative excitation to play a significant role in populating the (vibra-)rotational levels. For Einstein $A$-values in the range of 0.013 -- 0.019\,sec$^{-1}$ (see Table~\ref{Table:iden_NaCl}) and assuming collision rates within $v=0$ around $10^{-10}$\,cm$^{3}$sec$^{-1}$ \citep[based on the 
collisional rates of SiS,][]{Agundez2012A&A...543A..48A}, one gets a critical density $n_{\rm{crit}}$ around $2\times10^{8}$\,cm$^{-3}$. Using the output from the thermodynamical model for the wind of VY~CMa as presented by \citet{Decin2006A&A...456..549D}, this implies that there is only a small region for $r \la 9$\,\Rstar\ ($\sim$0.05\arcsec) where the gas number density is higher than the critical density and hence where LTE would prevail.  

\citet{Milam2007ApJ...668L.131M} determined the formation locus of gaseous sodium chloride to be close to the central star, between $\sim$5-40\,\Rstar, with a fractional abundance relative to H$_2$ of $5\times10^{-9}$. This is in accord with thermochemical equilibrium predictions, not taking phase changes into account, predicting that the gaseous NaCl abundance peaks around 500 to 1200\,K at densities $n\sim\,10^{11}$\,cm$^{-3}$ \citep[see Fig.~3 in][]{Milam2007ApJ...668L.131M}, with NaCl being the main sodium carrier in oxygen-rich environments\footnote{Based on the output from the thermodynamical model for the wind of VY~CMa as presented by \citet{Decin2006A&A...456..549D}, we note that a density of $n\sim\,10^{11}$\,cm$^{-3}$ is reached very close to the stellar atmosphere around 1.02\,\Rstar, where the gas kinetic temperature is well above 1200\,K, i.e.\ around 2700\,K.}. For temperatures above 1500\,K, the fractional abundance of gaseous NaCl is below $10^{-12}$. The confined distribution of gaseous sodium chloride, the fact that the observed velocities do not reach the terminal velocity, and realising the refractory nature of sodium chloride hint toward the fact that NaCl condenses easily onto grains. Thermochemical calculations including gas chemistry and the formation of possible liquid and solid condensates predict that sodalite, Na$_4$[AlSiO$_4$]$_3$Cl, is the major silicated chlorine containing condensate in an oxygen-rich environment, while halite, crystalline NaCl salt, might be formed in a carbon-rich wind \citep{Lodders1999IAUS..191..279L}. However, the fact that the ALMA data prove the existence of gaseous NaCl at distances far beyond the major dust condensation region, specifically around 1.22\arcsec\ ($\sim$220\,\Rstar) in the `SW' extension, is an incentive to reconsider the different gas-grain chemical processes at work that might prevent all gaseous NaCl species to be condensed. 

 A simplified NaCl chemical network is developed in Sect.~\ref{Sec:condensation} with the aim 
to constrain the main chemical processes that control the gaseous and solid NaCl content. 
Potential geometrical structures as derived from the multi-component NaCl emission are 
discussed in Sect.~\ref{Sec:Morphology}.

\subsection{NaCl gas-grain chemistry} \label{Sec:condensation}

Our aim is to understand the cycling between gaseous and solid NaCl through the envelope of 
VY~CMa. Since many parameters such as the sticking coefficient, surface binding energy, and 
the surface density of adsorbing sites are not known, we have constructed a simplified 
chemical network that lets us control several parameters that directly influence the NaCl 
abundance profile. Analogous simplified networks are constructed to study, for instance, 
the water content in low-mass forming regions \citep{Schmalzl2014A&A...572A..81S}. The main 
chemical processes to be considered are accretion of gaseous NaCl onto grains, thermal 
desorption and photodesorption.

\subsubsection{Simplified NaCl network} \label{Sec:simplified_network}

The rate of accretion of gaseous NaCl onto existing dust grains is given by 
\begin{equation}
R_{\rm{ACC}} = S \sigma_d \langle v_X \rangle n_g(\rm{NaCl}) 
\label{Eq:acc}
\end{equation}
in units of [cm$^{-3}$\,s$^{-1}$], where $S$ is the sticking probability, $\sigma_d$ the 
total grain surface cross-section per unit volume in units of [cm$^2$\,cm$^{-3}$], 
$\langle v_X \rangle$ is the average thermal velocity $\sqrt{(8 k T_g/m)}$ in units of 
[cm s$^{-1}$]  with $T_g$ the gas kinetic temperature, $k$ the Boltzmann constant, $m$ 
the mass of gaseous NaCl (58.5 atomic mass units), and $n_g(\rm{NaCl})$ the number density 
of gaseous NaCl in units of cm$^{-3}$.

We can write the total grain surface cross section per unit volume as
\begin{equation}
\sigma_d = \int \left(\pi a^2\right)\, n_d(a,r)\, da\,,
\end{equation}
with $n_d(a,r)$ being the grain volume density for a grain with radius $a$. We adopt a 
grain size distribution $n_d(a,r) da = K\, a^{-3.5}\, n_{\rm{H}}(r)\, da$, with a minimum 
grain size, $a_{\rm{min}}$, of 0.005\,$\mu$m and a maximum grain size, $a_{\rm{max}}$, of 
0.25\,$\mu$m. The slope of $-3.5$ is typical for interstellar grains 
\citep{Mathis1977ApJ...217..425M}. $n_{\rm{H}}$ is the total hydrogen number density in 
units of [cm$^{-3}$] and $K$ represents an abundance scaling factor giving the number of 
dust particles in units of particles per H atom. For the interstellar medium, $K$ is 
estimated to be around $7.9 \times 10^{-26}$cm$^{2.5}$/H for silicate grains 
\citep{Draine1984ApJ...285...89D}. For a dust-to-gas mass ratio of 0.002 \citep{Decin2006A&A...456..549D}, 
$K$ is $\sim$2.8$\times10^{-26}$\,cm$^{2.5}$ per H atom (see App.~\ref{App:additional_eq}). 
For the assumed grain size distribution, the total grain surface cross section $\sigma_d$ 
can hence be written as $\sigma_d\!\sim\!2500\, \pi\, K\, n_{\rm{H}}(r)$ or $\sigma_d\!\sim\!2.2 \times 10^{-22}\, n_{\rm{H}}(r)$.

The number density of gaseous NaCl, $n_g(\rm{NaCl})$, is written as 
\begin{equation}
n_g(\rm{NaCl})(r)=X(\rm{NaCl})\,n_{\rm{H}}(r)\,.
\nonumber
\end{equation}
 Assuming solar elemental abundance of chlorine ($\sim$3.2$\times10^{-7}$) and sodium 
 ($\sim$2.1$\times10^{-6}$) the abundance of NaCl, $X(\rm{NaCl})$, is controlled by the 
 chlorine content. 

The rate of thermal desorption is given by \citep{Tielens1982A&A...114..245T}
\begin{equation}
R_{\rm{TD}} = n_s(\rm{NaCl})\, \nu_0\, \exp(-E_{b}/k\,T_d) 
\label{Eq:des}
\end{equation}
in units of [cm$^{-3}$\,s$^{-1}$], where $n_s(\rm{NaCl})$ is the number density of NaCl 
molecules on the grains per cm$^3$, $\nu_0$ a characteristic vibration frequency in units 
of [s$^{-1}$], $E_{b}$ the surface binding energy, and $T_d$ the dust grain temperature.

To get the characteristic vibration frequency, $\nu_0$, we adopt the harmonic oscillator 
approach \citep{Hasegawa1992ApJS...82..167H}
\begin{equation}
\nu_0 = \sqrt{2\,N_s\,E_b/\pi^2\,m}\,,
\end{equation}
where $N_s$ is the surface density of adsorbing sites on the grains and $m$ is the mass of 
the adsorbed NaCl particle. The number of adsorbing sites per unit area, $N_s$, is assumed 
to be $\sim$1.5$\times 10^{-15}$\,cm$^{-2}$ \citep{Hasegawa1992ApJS...82..167H}. Hence $\nu_0$ 
is given by
\begin{equation}
\nu_0 = 2.07 \times 10^{12} \sqrt{\frac{T_b}{10000}}\,,
\end{equation}
with $T_b$ the surface binding temperature in Kelvin. This is in good accord with the 
observed frequency of 336\,cm$^{-1}$ (or $1\times10^{13}$\,s$^{-1}$) listed by 
\citet{Martin1983PhR....95..167M} or the value of 380\,cm$^{-1}$ (or $1.13\times10^{13}$\,s$^{-1}$) 
calculated using the Born-Mayer lattice theory by \citet{Rittner1951JChPh..19.1030R}.

One has to realize that even if all Cl is in NaCl and if all NaCl molecules start off in 
the gas and accretes onto already formed dust grains, NaCl will only occupy at most 25\% 
of the available surface sites, i.e.\ NaCl is in the sub-monolayer regime. The reason for that is that the number of sites per unit 
volume is given by $4\, \sigma_d\, N_s$ or $1.3 \times 10^{-6}\,n_{\rm{H}}(r)$ with the 
cosmic abundance of Cl being $3.2 \times 10^{-7}\,n_{\rm{H}}(r)$.  As shown 
by \citet{Tielens1982A&A...114..245T} quantum mechanical tunneling and thermal 
hopping of the adsorbed species on the grain surface are not relevant for NaCl.

For sub-monolayer coverage NaCl can photodesorb from the grain surface through far-ultraviolet (FUV) photons at a rate
\begin{equation}
R_{\rm{PD}} = (F_{\rm{FUV}} Y_{\rm{PD}}/4N_s)n_s(\rm{NaCl}),
\end{equation}
in units of [cm$^{-3}$\,s$^{-1}$]. $F_{\rm{FUV}}$ is the flux of FUV photons reaching the 
grain and $Y_{\rm{PD}}$ the photodesorption yield. 

The photodesorption yield $Y_{\rm{PD}}$ is difficult to assess. For grains for which water ice is the dominant mantle component, water molecules are photodesorbed either through the excess energy caused by the recombination of H and OH radicals following
direct photodissociation or via the 'kick-out' mechanism where an excited H atom ejects an H$_2$O 
molecule upon collision \citep{Arasa2015A&A...575A.121A}.  In VY~CMa \citet{Shenoy2013AJ....146...90S} derive a grain 
temperature of $\sim$ 130--170\,K at the `SW' clump, making it marginal whether ice mantles 
can be present - thermal desorption of water occurs at 160\,K \citep{Collings2004MNRAS.354.1133C}.
If no water ice is present, the yield is likely to be much less than $3\times10^{-3}$ which is
a typical value for molecules in water ice \citep{Oberg2009A&A...504..891O, Oberg2009ApJ...693.1209O, Westley1995P&SS...43.1311W}

Although it is commonly thought that the internal FUV radiation field of RSG stars is negligible, 
this is not necessarily the case for material sitting close to the stellar photosphere. To calculate the
flux at the `SW' clump, we use an effective stellar temperature, T$_{\star}$, of 3480\,K, the photodissociation
cross-sections of NaCl measured by \citet{Silver1986JChPh..84.4378S}, and a distance of 250\,R$_{\star}$ to the `SW' clump. 
For wavelengths shorter than 2910\,\AA, the threshold wavelength for NaCl photodissociation, 
we find that the total FUV flux is roughly 10$^4$ times larger than the general ISM flux \citep{Draine1978ApJS...36..595D}
and calculate the unshielded photodissociation rate to be
3.2$\times10^{-5}$ s$^{-1}$, which is a lower limit to the rate given that the cross-section
has not been measured below 1890\,\AA\ and some 10$^4$ times larger than the unshielded ISM
rate \citep{vanDishoeck1988ASSL..146...49V}.

The change of the gaseous and solid NaCl number densities is then given by
\begin{equation}
\frac{dn_g(\rm{NaCl})}{dt} = -\frac{dn_s(\rm{NaCl})}{dt}  =  - R_{\rm{ACC}} + R_{\rm{TD}} + R_{\rm{PD}}.
\label{Eq:rate_eqn}
\end{equation}
and we make the assumption that
\begin{equation}
 X(\rm{Cl})n_{\rm{H}}(r) = n_g(\rm{NaCl}) + n_s(\rm{NaCl})
 \label{Eq:cons_NaCl}
\end{equation}
at any time $t$ with $X(\rm{Cl})$ the cosmic abundance of chlorine relative to hydrogen.

\subsubsection{Chemical network model results} \label{Sec:results_chem_network}


\begin{figure*}[htp]
\begin{minipage}[t]{.3\textwidth}
        \centerline{\resizebox{\textwidth}{!}{\includegraphics[angle=270]{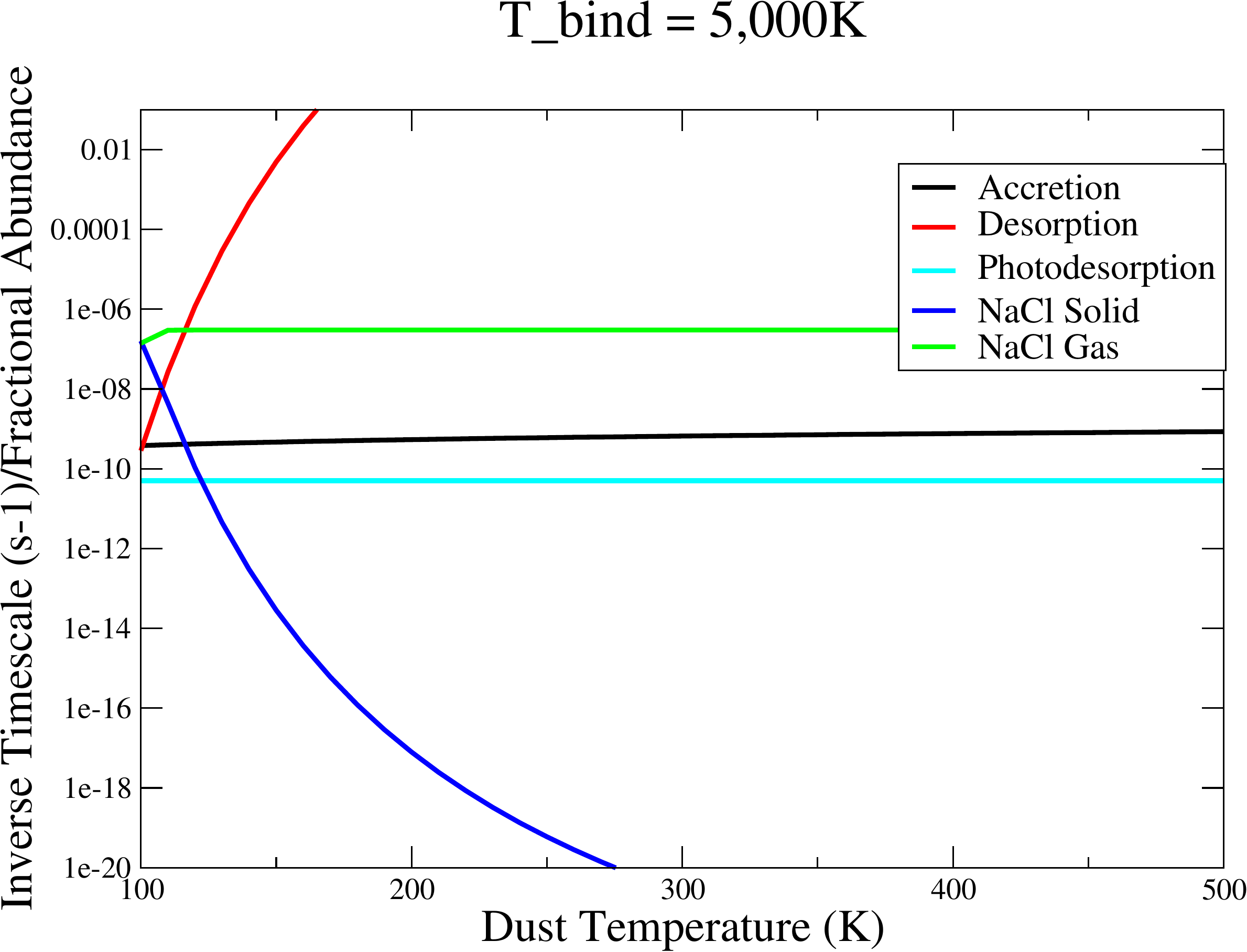}}}
    \end{minipage}
    \hfill
\begin{minipage}[t]{.3\textwidth}
        \centerline{\resizebox{\textwidth}{!}{\includegraphics[angle=270]{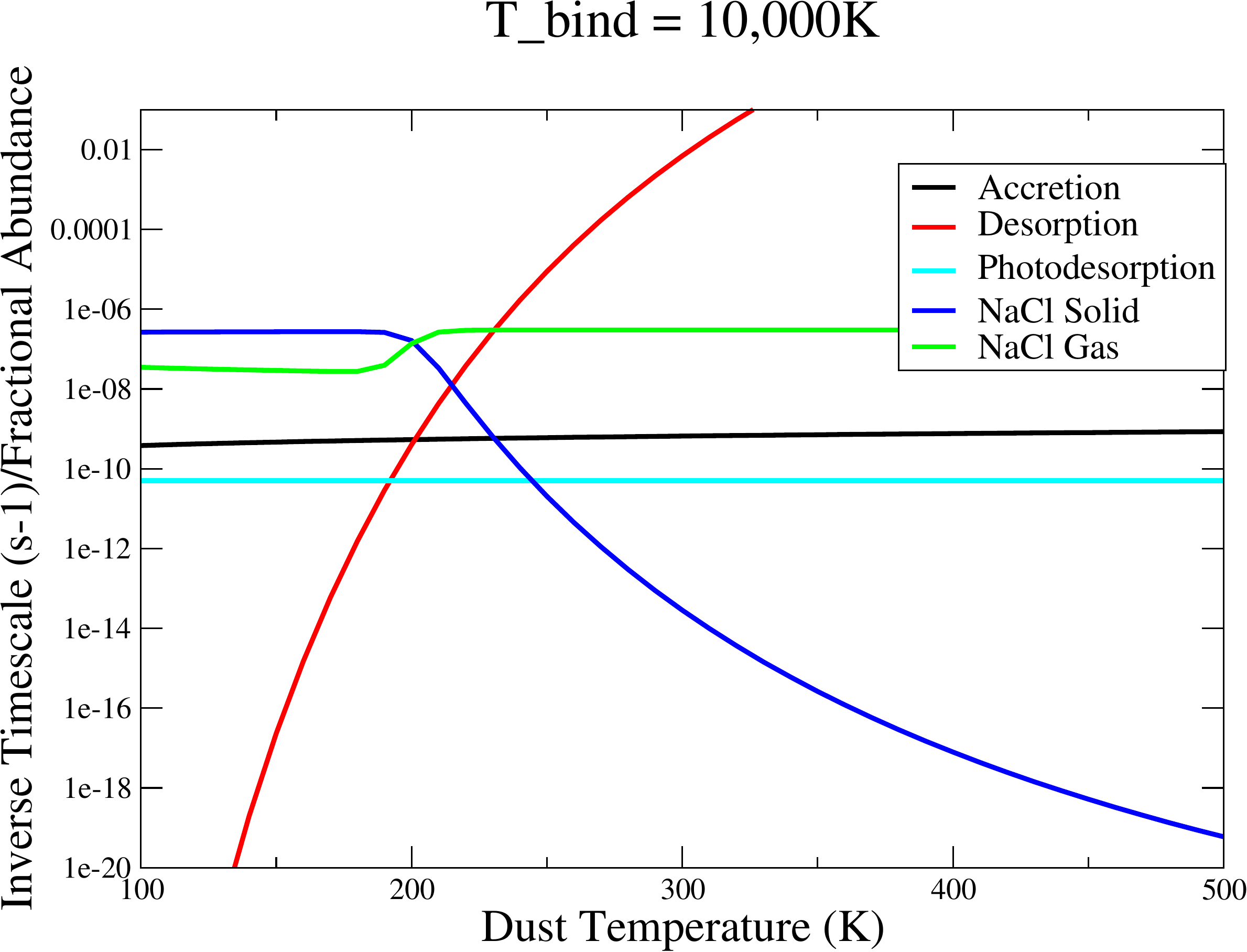}}}
    \end{minipage}
    \hfill
\begin{minipage}[t]{.3\textwidth}
        \centerline{\resizebox{\textwidth}{!}{\includegraphics[angle=270]{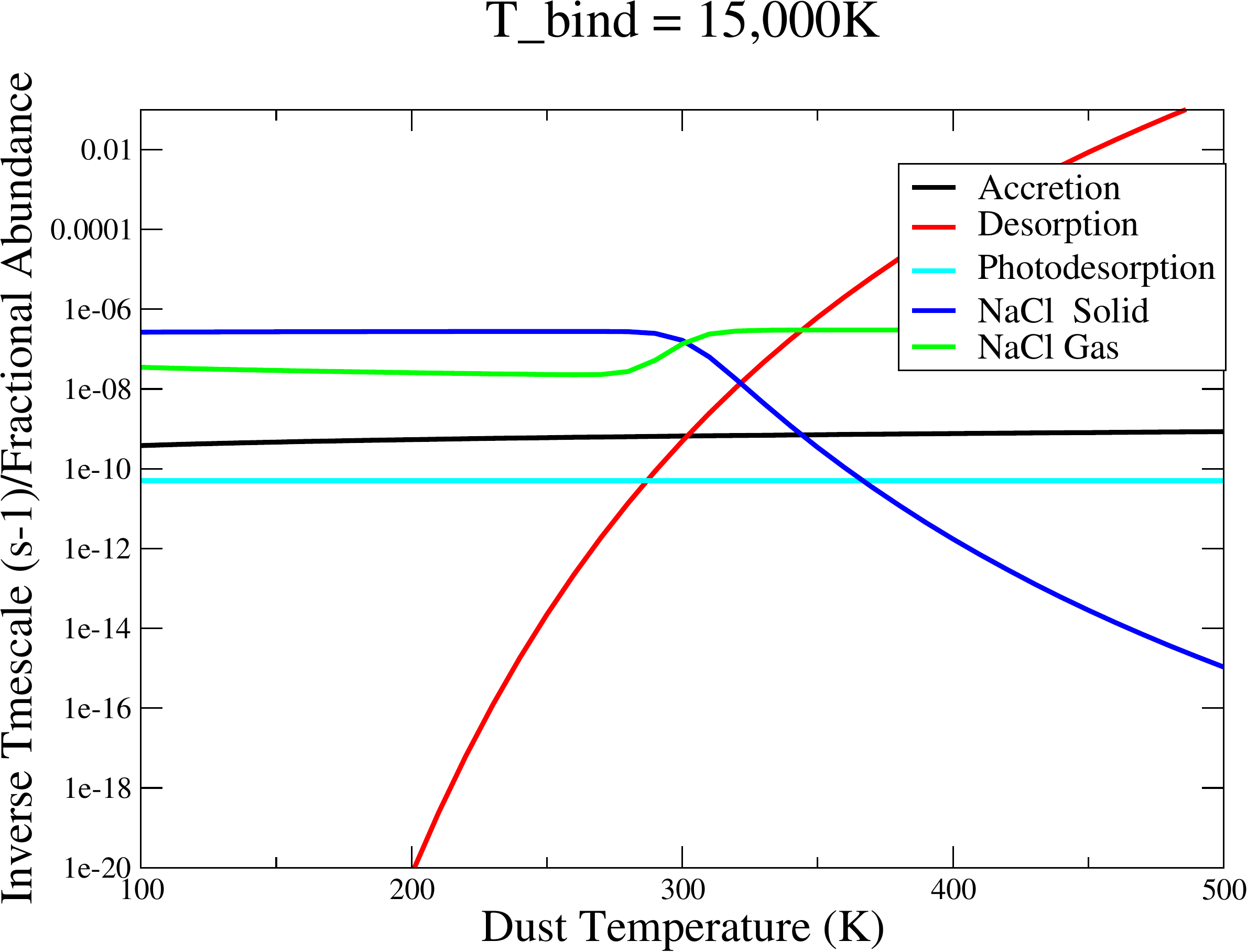}}}
    \end{minipage}
    \hfill
\caption{Steady state solutions for the gaseous and solid NaCl abundances (relative to the hydrogen number density) 
as a function of the dust temperature, $T_d$, for the case in which accretion, thermal 
desorption and photodesorption  are included in the 
chemical network. The different panels are for different 
values of the binding temperature (5\,000, 10\,000, 15\,000\,K) at a hydrogen number density 
of $1\times10^8$\,cm$^{-3}$. Each panel also shows the inverse timescale for accretion, 
thermal desorption and photodesorption in units of s$^{-1}$, calculated with $F_{\rm{FUV}}Y_{\rm{PD}} = 3 \times 10^5$.}
\label{Fig:steady_state}
\end{figure*}

\begin{figure*}[htp]
\begin{minipage}[t]{.3\textwidth}
        \centerline{\resizebox{\textwidth}{!}{\includegraphics[angle=270]{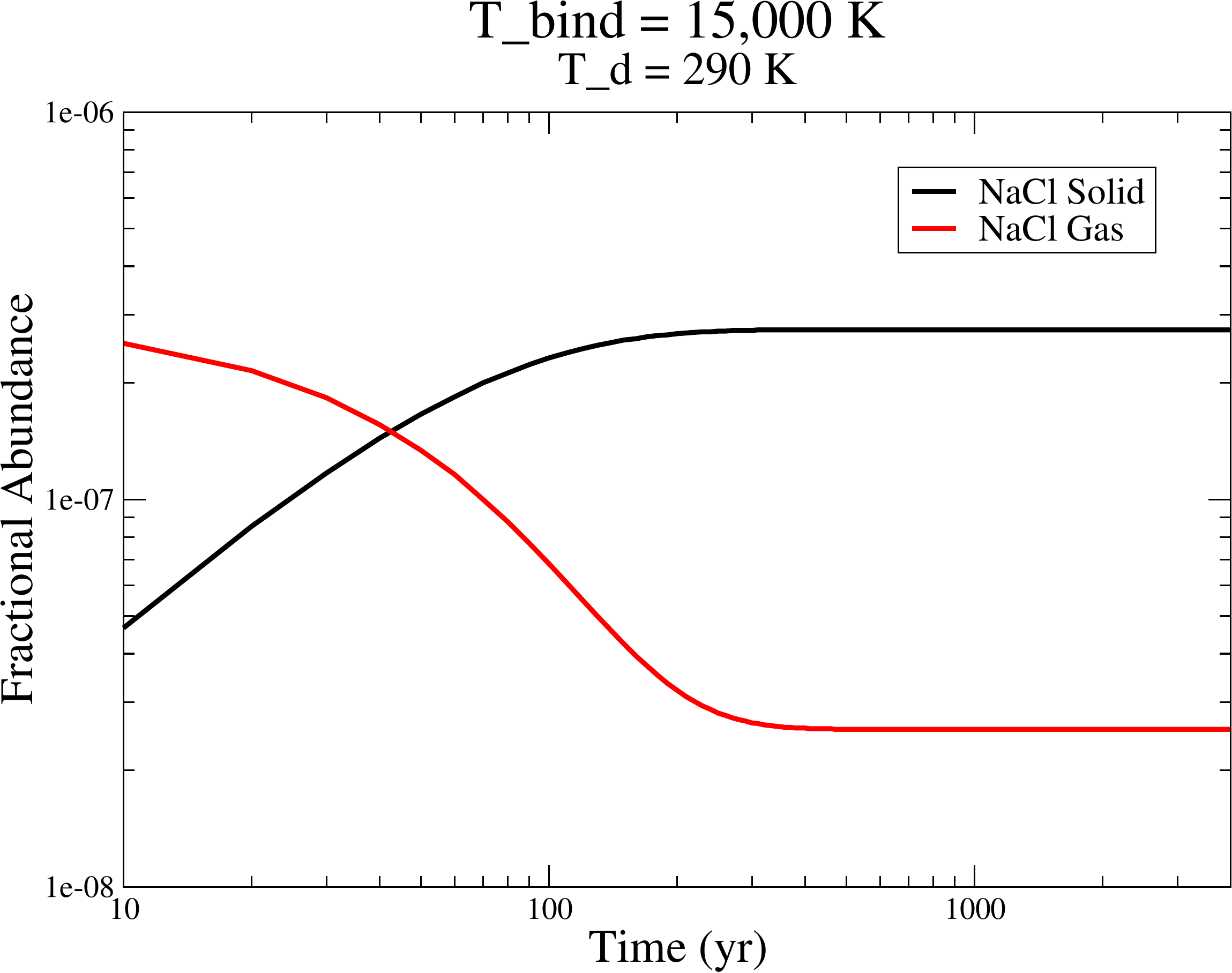}}}
    \end{minipage}
    \hfill
\begin{minipage}[t]{.3\textwidth}
        \centerline{\resizebox{\textwidth}{!}{\includegraphics[angle=270]{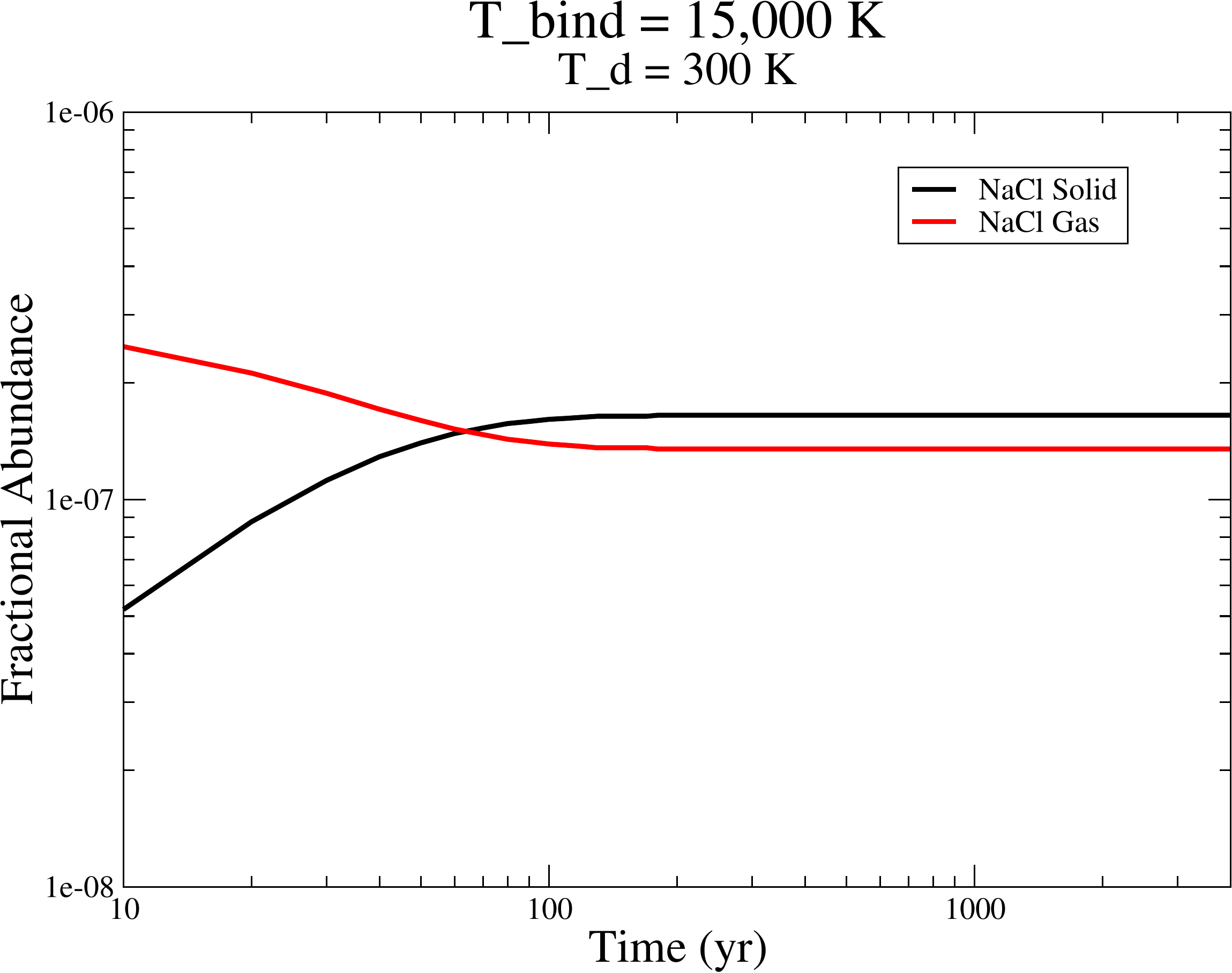}}}
    \end{minipage}
    \hfill
\begin{minipage}[t]{.3\textwidth}
        \centerline{\resizebox{\textwidth}{!}{\includegraphics[angle=270]{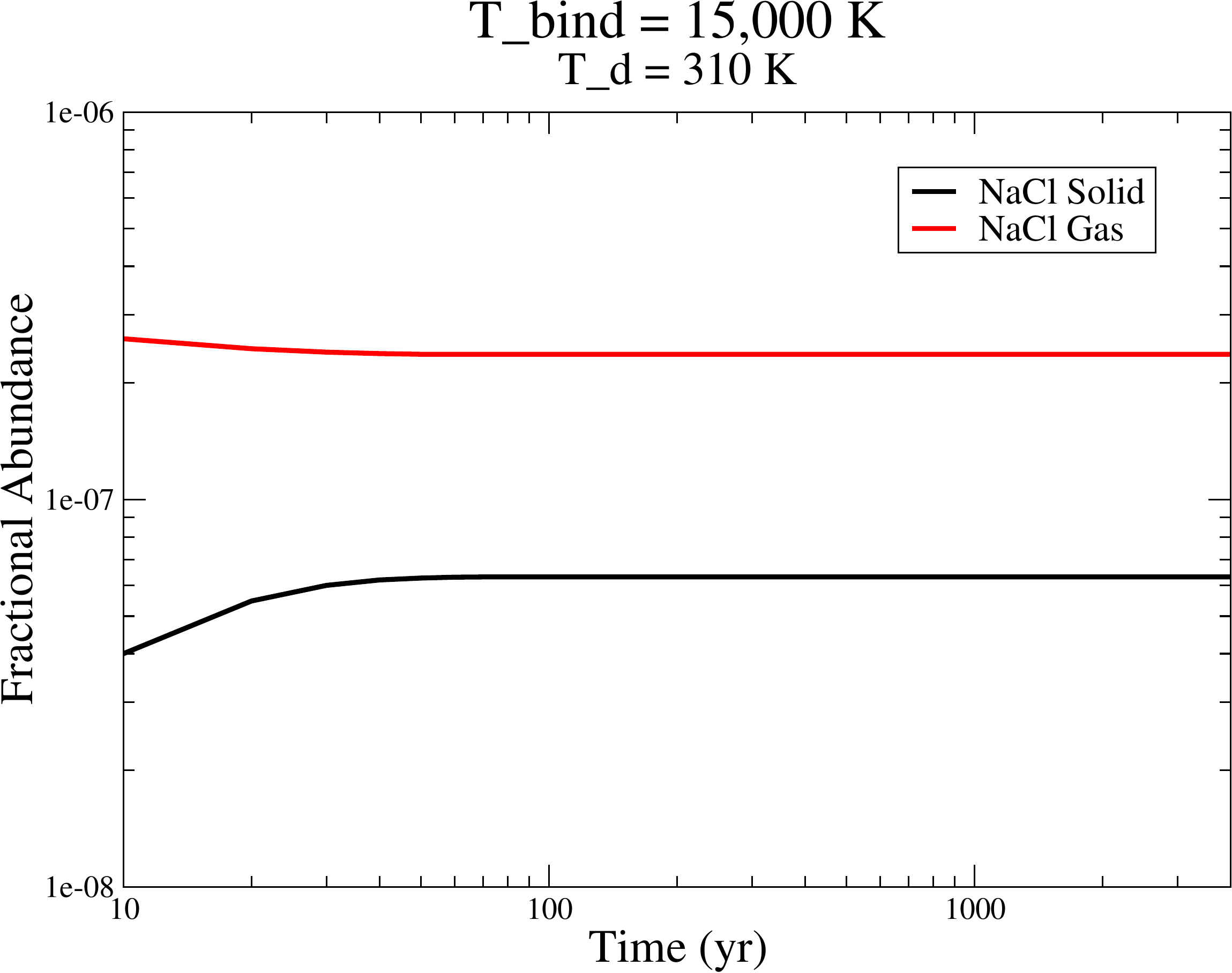}}}
    \end{minipage}
    \hfill
\caption{Fractional abundances of gaseous and solid NaCl in function of time for a binding temperature of 
15\,000\,K at three different values of the dust temperature $T_d$ (290, 300, 310\,K). }
\label{Fig:sol}
\end{figure*}

The value for the surface binding energy, $T_b$, is not known. It depends on the polarizability 
of NaCl and its ability to form specific bonds \citep{Tielens1982A&A...114..245T} to the surface. In the case 
of H$_2$O with its strong dipole moment and its ability to from hydrogen bonds, the surface 
binding energy on an ice surface has been calculated to be about 4000\,K \citep{Hale1981JChPh..75.1991H}. 
On the other hand, the surface binding energy of Na to a NaCl polycrystalline target is 
about 1.3\,eV (or $\sim$15\,000\,K). 
Considering this, we have decided to study the evolution of the NaCl abundance for three 
different values of the surface binding energies: 5\,000\,K, 10\,000\,K, and 15\,000\,K.

It is instructive to follow the abundance of gaseous and solid NaCl by solving Eq.~\ref{Eq:rate_eqn} 
for $n_s(\rm{NaCl})$.
Using conservation of NaCl at any time $t$ (Eq.~\ref{Eq:cons_NaCl}) and writing 
$\sigma_d(r)\,=\,2.2 \times 10^{-22}\,n_{\rm{H}}(r)\,=\,\sigma\,n_{\rm{H}}(r)$, the rate 
equation can be rewritten as 
\begin{equation}
\frac{dn_s(\rm{NaCl})}{dt} = a - b\,n_s(\rm{NaCl})\,,
\end{equation}
where $a\,=\,S \sigma n_H^2 v_X X(\rm{Cl})$ and $b\,=\,S \sigma n_H v_X + \nu_0 \exp(-T_b/T_d)
+ (F_{\rm{FUV}} Y_{\rm{PD}}/4N_s)$.
This has a simple time-dependent solution
\begin{equation}
n_s(\rm{NaCl},t) = \frac{a}{b}\left(1 - \exp(-bt)\right)\,.
\end{equation}
$n_s(\rm{NaCl},t=0) = 0$ and the time taken to reach steady state is a few times $1/b$. The 
steady-state solution is $n_s(\rm{NaCl},ss) = a/b$ and is relatively insensitive to the value of 
$S$ for $S \sim 1$.

It is difficult to determine the product $F_{\rm{FUV}}Y_{\rm{PD}}$ with any degree of certainty. 
The flux of FUV photons at the `SW' clump is very large if there is no extinction due to dust 
along the line-of-sight to the star.  If this really is the case, then NaCl is photodesorbed
completely on a very fast time-scale and the NaCl released from the dust will be
photodissociated rapidly, on a time-scale of a few days. In this case, NaCl is very unlikely 
to reform in the gas phase given that both Na and Cl are very minor elements. This scenario
is unlikely, however, as the dust condensation radius lies much closer to the star, on the 
order of 35--55 mas, see Sect.~\ref{Sect:comp_others}. Using our derived dust-to-gas mass ratio,
we calculate that the photosphere is hidden by some 10$^4$ magnitudes of visual extinction.
We note, however, that even 10 magnitudes of visual extinction is enough to reduce the 
unshielded FUV flux to interstellar values. Thus stellar FUV photons are unlikely to play 
a role in NaCl chemistry.

It is instructive, though, to consider a calculation for which the value of $F_{\rm{FUV}}Y_{\rm{PD}}$ = 3$\times10^{5}$, 
the typical interstellar value. Assuming a sticking coefficient, $S$, of 1, the steady-state solution for the gaseous and 
solid NaCl fractional abundances (with respect to the hydrogen number density) are shown 
in Fig.~\ref{Fig:steady_state} for different values of the binding energy. The hydrogen 
number density in Fig.~\ref{Fig:steady_state} is taken to be $1\times10^8$\,cm$^{-3}$. This 
value is derived from the equation of mass conservation
$\Mdot = 4\,\pi\,r^2\,\rho_g(r)\,v_g(r)$,
with $\rho_g$ the gas number density and $v_g(r)$ the gas velocity. Using a mass-loss 
rate $\Mdot$ of $1\times10^{-4}$\,\Msun\,yr$^{-1}$ and the gas velocity structure as derived 
from solving the momentum equation \citep[see][]{Decin2006A&A...456..549D}, a hydrogen 
number density of $1\times10^8$\,cm$^{-3}$ is reached at $\sim$10\,\Rstar, i.e.\ the dust 
condensation locus of interest to this study. Fig.~\ref{Fig:steady_state} also displays the 
inverse timescale for accretion and thermal desorption.  Fig.~\ref{Fig:steady_state} shows that 
photodesorption occurs at a slower rate than accretion and that only about 10\% of the NaCl 
abundance resides in the gas phase at low dust temperatures (relative to the surface binding temperature). For example, for a binding energy of 10\,000\,K, 
once the grain becomes warm enough, around 200\,K, all solid-state NaCl is thermally desorbed 
into the gas phase. It is clear from these simulations 
that the crucial factor determining the amount of solid versus gaseous NaCl is the ratio of 
the binding energy to the dust temperature, which appears in Eq.~\ref{Eq:des} for the thermal 
desorption. For a binding temperature of 5\,000\,K all NaCl will be gaseous unless the dust 
temperature is below $\sim$120\,K. A higher binding energy implies that the dust grains may 
attain higher temperatures before thermal desorption starts to dominate over the accretion rate. 
The turn-over point is around $T_b/T_d \sim 50$ (see also Fig.~\ref{Fig:sol}). Close to this 
value, steady state solutions occur on timescales of 50-100\,yr as shown in Fig.~\ref{Fig:sol}.

At the dust condensation locus of $\sim$10\,\Rstar, the grain temperatures might be in the 
range of 500-1000\,K, so that all NaCl will be in gaseous form independent of the assumed 
binding energy. This is in line with the ALMA observations presented in this paper. A steady 
state solution of $\sim$100\,yr for a velocity of $\sim$15\,km\,s$^{-1}$ (as derived from the 
ALMA NaCl data) yields a travel distance of 30\,\Rstar\ (or 160\,mas). When the grain temperatures 
becomes lower than 300\,K, solid NaCl starts to form from accretion onto existing grains. 
Whenever $T_d<100$\,K, another chemical process than thermal desorption needs to be invoked for 
NaCl to remain (partly) in gaseous form. The ALMA data clearly show gaseous NaCl emission 
around 1.22\arcsec\ in the south-western extension, at a place where also scattering by dust 
grains peaks (see Fig.~\ref{Fig:NaCl_312_25km_HST}). Assuming only direct stellar radiation, 
\citet{Shenoy2013AJ....146...90S} derived a radiative equilibrium temperature for the grains 
at that distance of $\sim$130-170\,K. If the grains indeed have that temperature at 1.22\arcsec, 
this would imply that either the binding temperature is around 5\,000\,K so that NaCl stays gaseous 
that far away from the star or that, in the case of higher values for the binding temperature, 
another process is important. One possibility is shock-induced sputtering driven 
by localised mass ejections (see Sect.~\ref{Sec:localized}. Such a process could lead, for example, to the 
different spatial distributions of refractory molecules, such as observed for NaCl and TiO$_2$, since their 
sputtering rates likely depend on their binding energies.

\subsection{Morphology} \label{Sec:Morphology}

The NaCl emission does not peak at the stellar position `VY', which can be explained by the 
presence of a density cavity close to the star, where the temperature is too high for 
gaseous NaCl to exist in high enough amounts (see previous sections). The morphology 
in the channel maps is clearly inconsistent with a spherical, uniformly expanding, envelope, 
which would exhibit a circular or ring morphology with decreasing radius for increasing velocity 
difference w.r.t.\ systematic velocity. Previous studies hint toward a preferred axis of symmetry 
oriented northeast-southwest at a position angle of $\sim$50\deg\ (see Sect.~\ref{Sec:introduction}). 
The central waist in the integrated NaCl intensity map aligns with the central waist seen in the dust 
continuum map and, as seen in Fig.~\ref{Fig:NaCl_312_HST} and Fig.~\ref{Fig:NaCl_312_25km_HST}, the 
ALMA data also roughly support the idea of a preferred axis oriented northeast-southwest.

Deducing the morphology of the circumstellar environment of VY~CMa is not a straightforward task. 
However, the ALMA NaCl data seem to favour two types of geometries: (1)~an axisymmetric geometry, 
be it a disk, a bipolar outflow, or in general a kind of toroidal overdensity or (2)~localized mass 
ejections. We consider the merits of both geometries regarding the morphology seen in the ALMA data 
and other optical, infrared and maser data published in the literature.

\subsubsection{Axisymmetric geometry}\label{Sec:equatorially_enhanced}

Within 1\arcsec\ of the central star, the ALMA NaCl data suggest an axial symmetry with a 
preferred axis oriented NE/SW at a PA of $\sim$50\deg\ (defined North-East), showing both 
a bright NaCl and dust continuum peak to either side of the symmetry axis. This kind of 
spatial-kinematic structure can be interpreted as an equatorially enhanced density structure 
with the polar axis, hence lower density region, along the NE/SW axis at PA of $\sim$50\deg. 

Different studies (see Sect.~\ref{Sec:introduction} and below) support the axisymmetry 
revealed by the ALMA data. \citet{Herbig1970ApJ...162..557H} invoked a flared disk or ring 
in latitudes 10\deg\ to 30\deg\  and seen almost edge-on to explain the infrared energy 
distributions. Using optical and near- and mid-infrared data \citet{Smith2001AJ....121.1111S} 
gave support to that interpretation and qualitatively interpreted the morphological structure 
as being caused by an equatorial density enhancement with a bipolar axis projected roughly 
northeast-southwest, and with the southwest lobe closer to us. The lack of optical and 
near-infrared emission in the northeast region of their data could be explained by high 
extinction along the line-of-sight and inefficient backscattering of the light by the dust 
particles. The SMA data of $^{12}$CO and $^{13}$CO at a spatial resolution of $\sim$2\arcsec\ 
were modeled by \citet{Muller2007ApJ...656.1109M} using a dense, compact and dusty central 
component, embedded in a more diffuse extended envelope and with a high-velocity bipolar 
outflow oriented in the east-west direction with a wide opening angle ($\sim$120\deg) viewed 
close to the line-of-sight ($i=15$\deg). The inner radius of the bipolar lobes was determined 
at $\sim$0.6\arcsec\ and the velocity in the bipolar lobes increased linearly from 
15\,km\,s$^{-1}$ to 45\,km\,s$^{-1}$ at the outer radius. An enhanced density shell 
within the bipolar lobes was needed to fit the observed CO data.

\begin{figure*}
   \begin{minipage}[t]{.33\textwidth}
        \centerline{\resizebox{\textwidth}{!}{\includegraphics{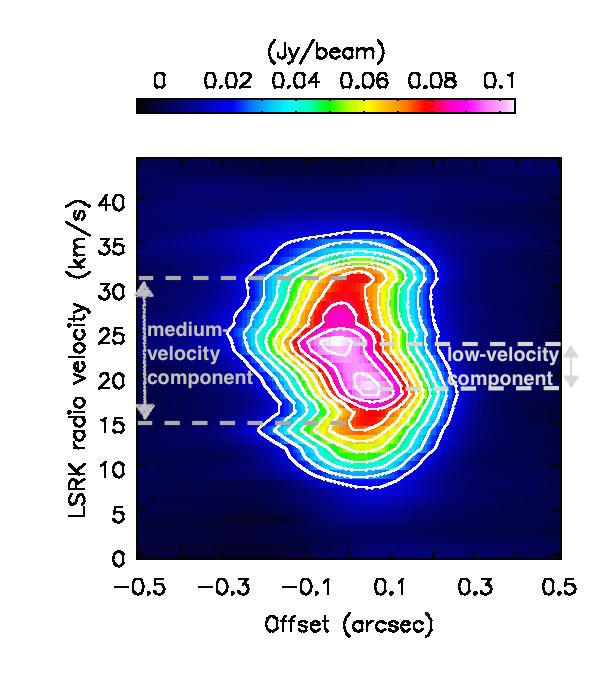}}}
    \end{minipage}
    \hfill
    \begin{minipage}[t]{.33\textwidth}
        \centerline{\resizebox{\textwidth}{!}{\includegraphics{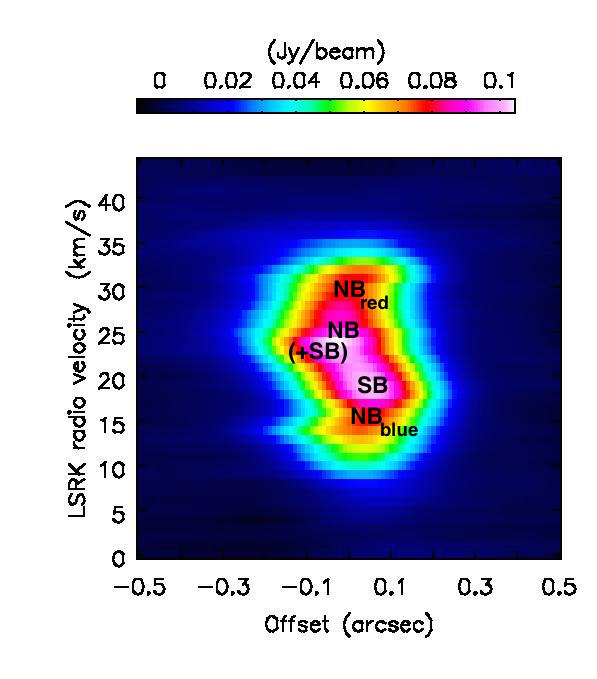}}}
    \end{minipage}
    \begin{minipage}[t]{.33\textwidth}
        \centerline{\resizebox{\textwidth}{!}{\includegraphics{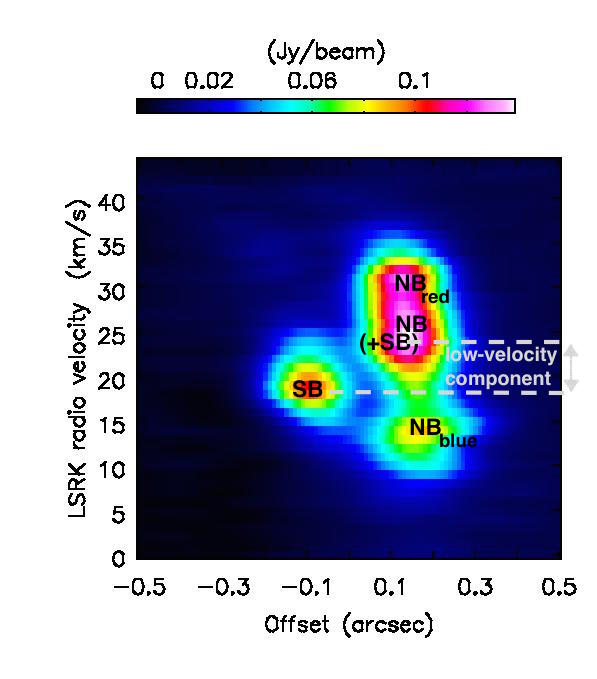}}}
    \end{minipage}
\caption{Position-velocity diagram of the NaCl 312\,GHz for an axis length of 1\arcsec\ and an averaging width of 650\,mas centred at $\alpha$\,=\,07h22m58.332s and $\delta$\,=\,$-$25\deg46\arcmin03\farcs053. In the left and middle panel, the slit has a PA of 50\deg\ (i.e.\ along the presumed polar/rotation axis; with north-east positions having negative offsets and south-west positions positive offsets), in the right panel the slit PA is 140\deg\ (i.e.\ along the presumed equatorial plane; with south-east positions having negative offsets and north-west positions positive offsets). In the left and right panel the different velocity components are indicated, while in the middle and right panel the main contributors to the different intensity peaks are indicated. The contours in the left panel are at $[1,2,...10]\times0.01$\,Jy/beam.}
 \label{Fig:NaCl_PV}
\end{figure*}

Two position-velocity (PV) diagrams of the NaCl 312\,GHz ALMA observations are shown in Fig.~\ref{Fig:NaCl_PV}; the figure to the left (PV1) gives the PV diagram for a slit oriented along the presumed polar/rotation axis at a PA of 50\deg\ to the north-east and the figure to the right shows the PV diagram (PV2) for a slit perpendicular to the polar/rotation axis, i.e.\ along the presumed equatorial plane.

PV1 diagram is characterized by a $S$-type signature. The northern clump is the main contributor to the bright peak around 24\,km\,s$^{-1}$ (see middle panel in Fig.~\ref{Fig:NaCl_PV}), while both the northern (NB$_{\rm{blue}}$) and southern clump contribute to the blue-shifted peak around 19\,km\,s$^{-1}$. The blue-shifted extension around 14\,km\,s$^{-1}$ arises from NB$_{\rm{blue}}$, while NB$_{\rm{red}}$ explains the extension toward velocities of $\sim$35\,km\,s$^{-1}$.
It is tempting to interpret the two horizontal ridges at $\Delta v \sim \pm 2.5$\,km\,s$^{-1}$ from the systemic velocity as arising from a disk (or in general a density enhanced equatorial region) inclined to the line-of-sight (hereafter referred to as the low-velocity component; see left panel in Fig.\ref{Fig:NaCl_PV}). The disk has a projected velocity around 2.5\,km\,s$^{-1}$. The blue-shifted part of the disk is to the south-west, and the red-shifted part to the north-east. There is also a more or less vertical ridge extending to $\pm$15\,km\,s$^{-1}$ w.r.t.\ the systemic velocity, with 2 other horizontal ridges at $\Delta v \sim \pm 9$\,km\,s$^{-1}$ from the systemic velocity (i.e.\ at 14 and 31\,km\,s$^{-1}$). This medium-velocity component\footnote{Wind speeds up to $\sim$45\,km\,s$^{-1}$ are present in the nebula surrounding VY~CMa \citep{Decin2006A&A...456..549D, Smith2004MNRAS.349L..31S, Muller2007ApJ...656.1109M}. We therefore opt to use `low velocity' for $v<5$\,km\,s$^{-1}$ and `medium velocity' for $5<v<15$\,km\,s$^{-1}$.} is only seen in the northern clump, with the blue-shifted part slightly more to the north-east than the red-shifted part, i.e.\ an orientation opposite to the low-velocity disk. This tentatively hints towards an interpretation that the medium-velocity component is a biconical polar outflow, with wide opening angle, perpendicular to the equatorial disk plane. An analogous morphology  was also deduced for the carbon star V~Hya by \citep{Hirano2004ApJ...616L..43H} and the S-type AGB star $\pi^1$~Gru by \citet{Chiu2006ApJ...645..605C} from the analysis of $^{12}$CO J=2-1 SMA data, showing similar PV diagrams as PV1. With the intensity of `NB' being brighter around 31\,km\,s$^{-1}$ than at 14\,km\,s$^{-1}$, this suggests that the far-side of the bipolar outflow is situated to the north. This orientation supports the analysis by \citet{Smith2001AJ....121.1111S}, who suggested that the far side of the optical/near-infrared nebula is situated to the north-east, while the southwest lobe is closer to us. A geometry with an equatorial density enhancement roughly northwest-southeast and with the polar lower density region oriented along a NE/SW axis can also explain the morphology of the dust continuum as seen in the ALMA data and shown in Fig.~\ref{Fig:NaCl_312} \citep{OGorman2015A&A...573L...1O}. 

However, one then might wonder why the medium-velocity wind (aka bipolar outflow) is only seen in the northern clump (see also PV2), if the southwest lobe is closer to us. In the case of optically thin emission, it might be that the accumulation of optical thickness along the line sight is too low when reaching the southern clump and that the sensitivity of the present ALMA data is not high enough to trace these low intensities. Another suggestion is that for one reason or another (see Sect.~\ref{Sec:condensation}) gaseous NaCl condenses rapidly in the southern lobe and that its abundance is too low to be traced.

However, this axisymmetric geometry can not explain the full morphological picture of the envelope around VY~CMa. As clearly shown by the ALMA detection of NaCl emission in the `SW' clump  and as already mentioned in this paper, there are convincing indications of localised density structures that might be caused by randomized mass ejections from the ill-defined convective surface of the star. However, even in that case, the proposed lower density polar regions in the northeast-southwest direction might explain why we see the emission from the NaCl clump at 1.2\arcsec\ to the southwest: this clump, that might be caused by a higher density mass-loss event, can receive more illumination from the central star through the low density region which will enhance its excitation. This effect was also invoked by \citet{DeBeck2015} to explain the detection of TiO$_2$ in the `SW' clump.

\subsubsection{Localized mass ejections}\label{Sec:localized}

The present ALMA NaCl data can also be geometrically interpreted as showing evidence for 
localized mass ejections. Episodic mass loss events were already suggested by, e.g., \citet{Ziurys2007Natur.447.1094Z} and \citet{Smith2009AJ....137.3558S}; see Sect.~\ref{Sec:introduction}. As described in Sect.~\ref{Sec:spatial_distribution}, the flux density of the `NB' shows two local maxima, one around 14\,km\,s$^{-1}$ (`NB$_{\rm{blue}}$') 
and one around 24\,km\,s$^{-1}$ (`NB$_{\rm{red}}$'). The minimum around 18\,km\,s$^{-1}$ 
coincides with the maximum in flux density of the southern clump (`SB'). Careful inspection 
of the channel maps in Fig.~\ref{Fig:NaCl_312} also shows that (albeit barely spatially resolved) 
the morphology of `NB$_{\rm{blue}}$' has a more pronounced north-south extension than `NB$_{\rm{red}}$'. 
While the TiO$_2$ emission coinciding with the `SW' clump is blue-shifted \citep{DeBeck2015}, 
the NaCl channel maps clearly show that the main emission component in this case is slightly 
red-shifted.
The spatially distinct clumps at different projected velocities might hence be interpreted 
as tracers of different localized mass ejection events, and it is even plausible 
that `NB$_{\rm{blue}}$' marks an event different from `NB$_{\rm{red}}$'. Some of these 
events are heading away from the observer, other ones are approaching us. This type of 
3-dimensional geometry has already been suggested in the literature (see Sect.~\ref{Sec:introduction}). 
Combining visual and near-IR images and 4.6\,$\mu$m CO spectra \citet{Smith2009AJ....137.3558S} 
proposed a geometry where a halo is responsible for an ubiquitous 25\,km\,s$^{-1}$ outflow 
with potentially mild axisymmetry. Faster ($\sim$40\,km\,s$^{-1}$) dense CO cloudlets move 
through this halo and create a very asymmetric envelope. The NaCl clumps clearly trace much 
lower velocities, albeit their emission is detected far beyond the dust formation radius 
estimated to be around 10\,\Rstar\ \citep [or 55\,mas][]{Monnier1999ApJ...512..351M}. We 
can not exclude the coincidental fact that the different NaCl clumps have space vectors 
quite close to the plane of the sky. However, it might be that these clumps indeed trace 
lower velocity structures. We note that \citet{Kaminski2013A&A...551A.113K} also deduced 
a velocity around 15\,km\,s$^{-1}$ for the `SW' clump, but other molecules in the SMA 
survey which emit closer to the central target trace velocities larger than the ALMA 
NaCl `NB' and `SB' clumps. In the case that the NaCl clumps and CO cloudlets are part of 
the same physical structure, the interaction interface between the cloudlet and the 
surrounding material could enhance the gas temperature thus facilitating the excitation 
of both CO and NaCl.

In the case of localized mass ejections, each event will have more or less the same chemical content at the moment of its ejection and the same laws of thermodynamics and chemistry will apply to each of them. One then might wonder why the morphology as seen in NaCl is so different compared to what is traced by the ALMA TiO$_2$ channel maps presented by \citet{DeBeck2015}, a molecule that also has a high dipole moment (6.33 Debye) and with an expected abundance slightly higher than NaCl in the inner wind region \citep{Gobrecht2015arXiv150907613G}. We simply note here that the \textbf{rotational energy level diagram} diagram of NaCl is less complex than for TiO$_2$ and that even different TiO$_2$ lines with similar excitation energies show quite different channel maps. This leads us to propose that the different mass ejections have very short dynamical times and different molecules/excitations might (partly) trace different events where the local thermodynamical, radiative and chemical properties are different. The nature of the star led \citet{Smith2001AJ....121.1111S} to suggest that these ejection events could be of convective or magnetic origin, alike the Sun with its solar flares and eruptions.

There is, however, some similarity between the ALMA NaCl and TiO$_2$ emission, i.e.\ both molecules trace the `SW' clump out to $\sim$1.2\arcsec, a clump also detected in SiS, CS, NS, and H$_2$S in the SMA data of \citet{Kaminski2013A&A...551A.113K} and in dust scattered light out to 1.4\arcsec\ \citep{Smith2001AJ....121.1111S, Shenoy2013AJ....146...90S}. Some of these molecules are expected to be quite abundant (as SiS) and others not, and they have very different dipole moments (from 1 to 9\,Debye). This might indicate that they trace the same physical event along a northeast-southwest axis. The `SW' clump is potentially embedded within the lower density southwest bipolar lobe (see Sect.~\ref{Sec:equatorially_enhanced}) or it might be that the dust clump is a high mass-loss event leaving a lower density wake behind it. In both cases, the lower density facilitates the radiative excitation of the molecules by direct stellar light.  And while the nature of the bright `C' component in the ALMA dust continuum map is not yet clear \citep{OGorman2015A&A...573L...1O}, in retrospective it might be that this dust clump also swept up material while moving away from the star leaving a lower density region behind it that facilitates the excitation of NaCl seen in the `SB', this under the assumption that both are in the same direction w.r.t.\ the central star.

\section{Conclusions}\label{Sec:conclusions}

In this paper we have analysed gaseous NaCl in the inner wind of the luminous red supergiant 
VY~CMa. The high sensitivity and high spatial resolution capabilities of ALMA allow a detection 
of this molecule within 250\,\Rstar\ from the central star. Four NaCl lines are detected, 
being the $v=0,1$ $J=24-23$ and $J=25-24$ rotational lines in ALMA band 7  at a spatial 
resolution of $\sim0\farcs24\times0\farcs13$. The ALMA data permit the first spatial 
decomposition of the NaCl emission in up to four different spatial regions. We show that 
all NaCl lines show a bimodal spatial distribution, with a bright emission component to
the north of the central star and a slightly weaker emission component to the south at an 
average separation of 100\,mas (or $\sim$18\,\Rstar) and 190\,mas (or $\sim$34\,\Rstar), 
respectively. The rotational NaCl lines in the ground vibrational state show in addition 
an extra emission component at $\sim$1.22\arcsec\ to the southwest of the star. This 
southwest clump is already detected in other molecules (CS, SiS, NS, and H$_2$S) with the 
SMA  \citep{Kaminski2013A&A...551A.113K} and in TiO$_2$ with ALMA \citep{DeBeck2015} and is 
aligned with the southwest extension seen in dust scattered light \citep{Smith2001AJ....121.1111S} 
and  other near-infrared and polarimetric data \citep[e.g.][]{Monnier1999ApJ...512..351M, Jones2007AJ....133.2730J}. 

The NaCl lines are detected up to a velocity of 17\,km\,s$^{-1}$ w.r.t.\ the local 
standard of rest velocity, $v_{\rm{LSR}}$, of 22\,km\,s$^{-1}$. This is significantly lower 
than the terminal velocity, which is around 35\,km\,s$^{-1}$ as estimated from low-$J$ 
rotational CO lines. This low velocity was already noticed by \citet{Milam2007ApJ...668L.131M}, 
who explained this as being due to the very efficient condensation of NaCl, being a highly 
refractory species. However, the fact that we detect gaseous NaCl at distance of 1.2\arcsec, 
far beyond the main dust formation region estimated around 35-55\,mas ($\sim$10\,\Rstar), 
points toward a chemical process preventing all gaseous NaCl from nucleating. We construct a 
simplified chemical network aiming at understanding the cycling between gaseous and solid 
NaCl through the envelope of VY~CMa. We show that at the dust condensation locus where the grain temperatures can be above 300\,K, thermal 
desorption dominates the rate of accretion of gaseous NaCl onto existing dust grains. 
For grain temperatures lower than 100\,K, photodesorption is unlikely to be 
important and removal of NaCl from the dust is likely due to shock sputtering driven
by localized mass ejection events.


It is not straightforward to interpret the geometrical signatures in the inner wind. Based 
on the ALMA data, and previous optical, near-IR, and mid-IR data two scenarios are favoured. 
In the first scenario, the overall density structure in the inner wind is described by an 
equatorial density-enhanced region with the lower density bipolar regions oriented 
northeast-southwest at a position angle of $\sim$50\deg. The density-enhanced region, be 
it a disk, has a projected velocity of 2.5\,km\,s$^{-1}$. The far side of the bipolar 
lobes is situated to the north, in agreement with the non-detection of the optical reflection 
nebula to the northeast of the star by \citet{Smith2001AJ....121.1111S}. However, this picture 
alone can not explain all morphological signatures. The south-west clump at 1.2\arcsec\ and 
the presence of arcs and knots in optical images point toward the action of localised mass 
ejections creating higher density structures within the diffuse halo. The present data, 
obtained with only 16--20 ALMA antennas, can not prove the validity of both scenarios, 
albeit it seems likely that both play a role in shaping the wind of VY~CMa.

\begin{acknowledgements}
This paper makes use of the following ALMA data: ADS/JAO.ALMA2011.0.00011.SV. ALMA is a partnership of ESO (representing 
its member states), NSF (USA) and NINS (Japan), together with NRC 
(Canada) and NSC and ASIAA (Taiwan), in cooperation with the Republic of 
Chile. The Joint ALMA Observatory is operated by ESO, AUI/NRAO and NAOJ.
\end{acknowledgements}

\bibliographystyle{aa}
\bibliography{VYCMa_NaCl}

 \newpage
 \begin{appendix}
 \section{Channel maps of the NaCl v=0,1 J=25-24 lines} \label{App:extra_channel_maps}
\begin{figure*}[htp]
\sidecaption
\includegraphics[width=120mm]{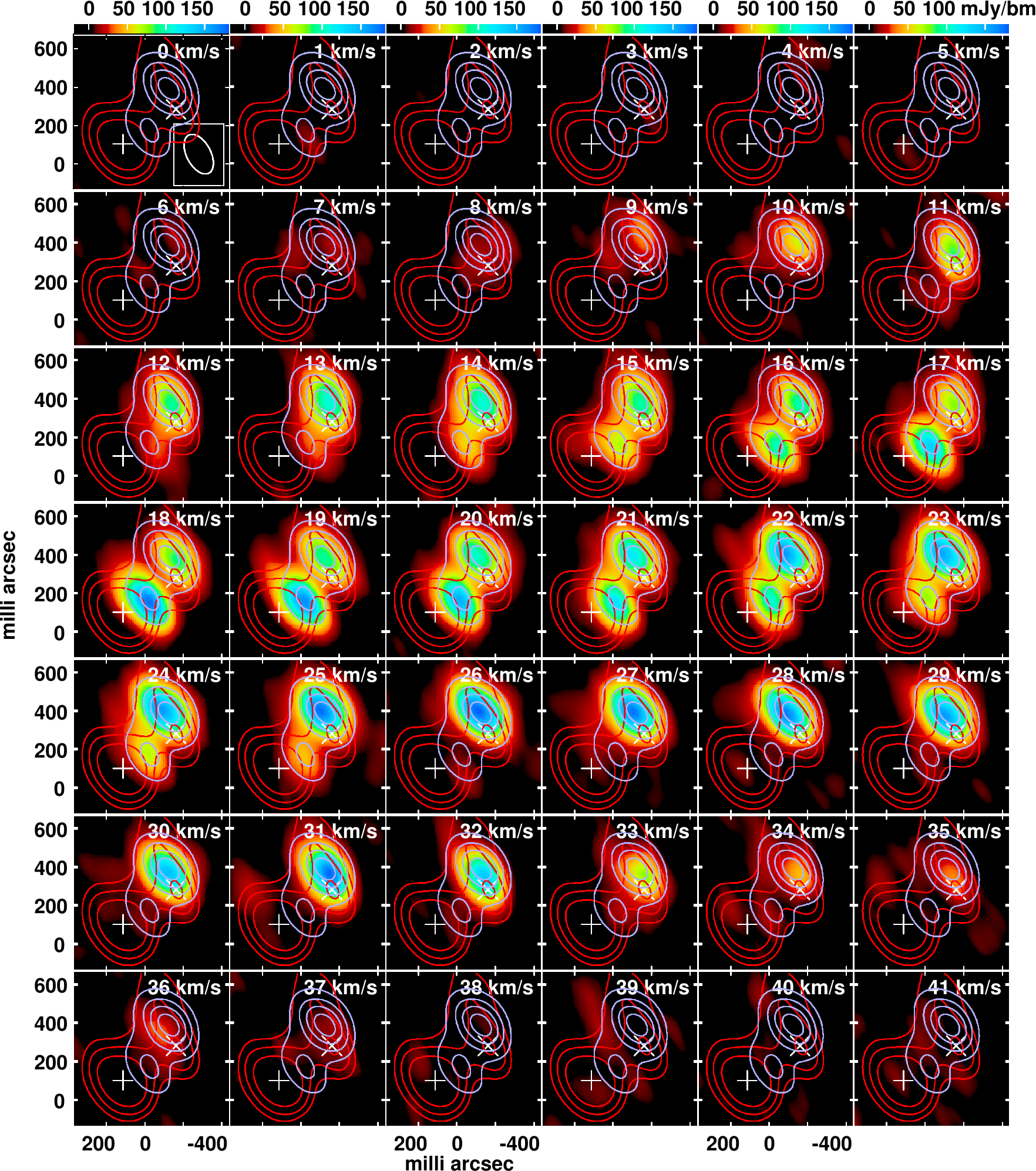}
\caption{Same as Fig.~\ref{Fig:NaCl_312} but for the NaCl v=1 J=25-24 line at 322\,GHz.}
\label{Fig:NaCl_322}
\end{figure*}

\begin{figure*}[htp]
\sidecaption
\includegraphics[width=120mm]{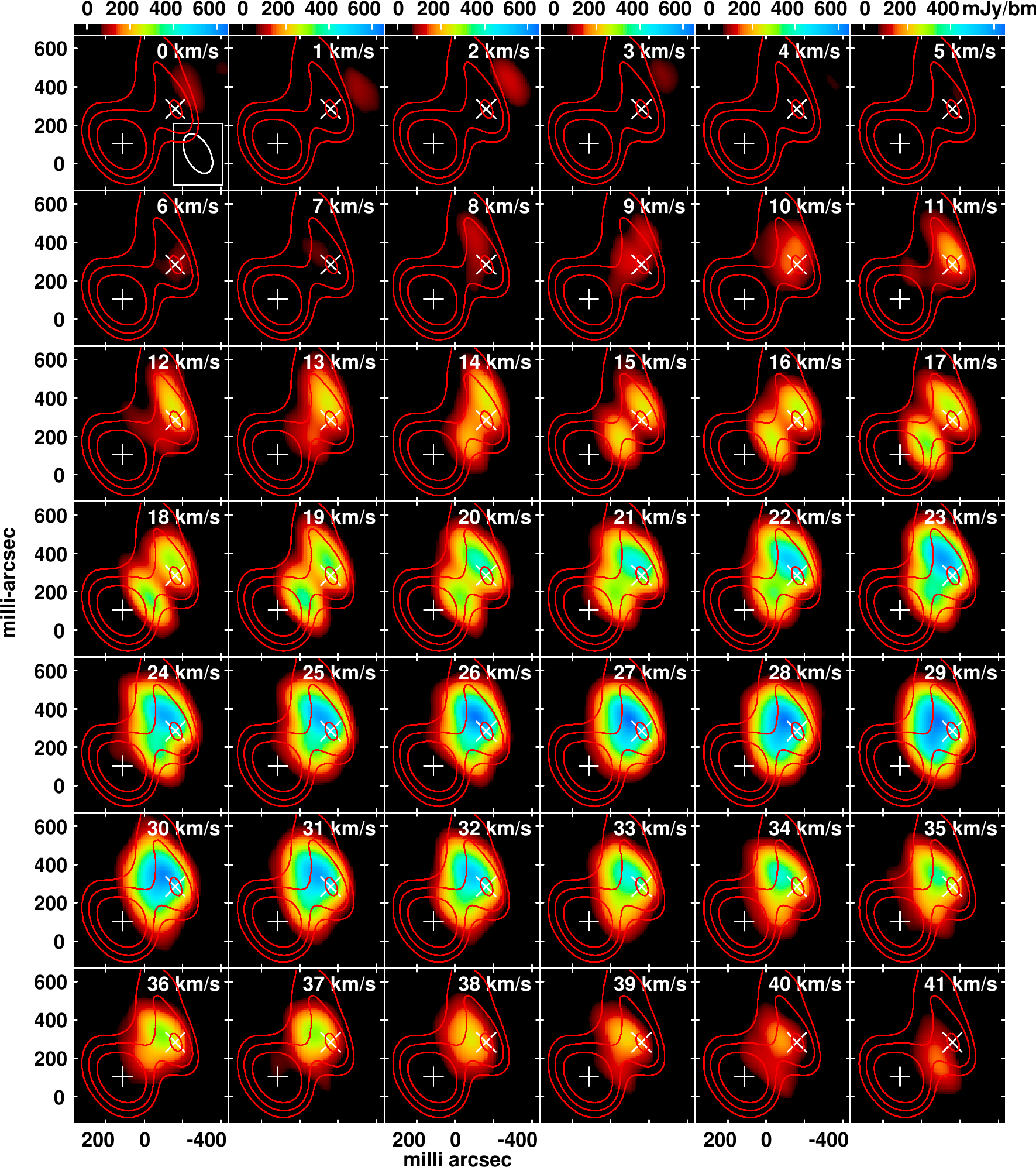}
\caption{Same as Fig.~\ref{Fig:NaCl_312} but for the NaCl v=0 J=25-24 line at 325\,GHz. 
Since this line is severely blended with the 325\,GHz H$_2$O maser line and the SiS v=1 
J=18-17 line the integrated line strength is not shown.}
\label{Fig:NaCl_325}
\end{figure*}

 Fig.~\ref{Fig:NaCl_322} and ~\ref{Fig:NaCl_325} show the channel maps for the NaCl v=0,1 J=25-24 emission lines.
 
\section{Additional equations}\label{App:additional_eq}

The dust-to-gas mass ratio, $\psi$, is defined as 
\begin{equation}
\psi = \frac{n_d\, m_d}{\rho_g}\,,
\end{equation}
with $n_d(a,r)$ being the grain volume density (see Sect.~\ref{Sec:condensation}), $m_d$ 
the mass of the grain being $4/3\, \pi\, a^3\, \rho_s$, with $\rho_s$ the specific density 
of the dust grain taken to 3.3\,g\,cm$^{-3}$ for silicate grains \citep{Draine1984ApJ...285...89D}, and $\rho_g$ the gas mass density 
being $n_{\rm{H}}(r)\,m_{\rm{H}}$, with $m_{\rm{H}}$ the mass of a hydrogen atom. Applying 
the same grain size distribution as described in Sect.~\ref{Sec:condensation}, it is easily 
shown that 
\begin{equation}
\psi = 7 \times 10^{22}\, K\,.
\end{equation}
For a dust-to-gas mass ratio of 0.002, this implies that $K$ is equal to $2.8 \times 10^{-26}$\,cm$^{2.5}$ per H atom.

The grain density $n_d(r) = \int n(a,r)\,da$ is then given by $2.2\times10^{15}\,K\,n_{\rm{H}}(r)$, 
which for the given value of $K$ results in $n_d(r) = 6\times10^{-11}\,n_{\rm{H}}(r)$.

\end{appendix}

\end{document}